\documentclass[aps,prl,twocolumn,showpacs,amsmath,amssymb,floatfix,notitlepage,groupedaddress]{revtex4-1}

\usepackage{color}
\usepackage{graphicx}
\usepackage{dcolumn}
\usepackage[utf8]{inputenc}
\usepackage{graphicx}
\usepackage{epstopdf}
\usepackage{amsthm}
\usepackage{amsmath}
\usepackage{empheq}
\usepackage{braket}
\usepackage{amssymb}
\usepackage[usenames,dvipsnames]{xcolor}
\usepackage{cases}
\usepackage{latexsym}
\usepackage[colorlinks=true,citecolor=Cerulean,linkcolor=RubineRed,urlcolor=Cerulean]{hyperref}
\usepackage[thinlines]{easytable}
\newcommand{\be}{\begin{equation}}
\newcommand{\ee}{\end{equation}}
\newcommand{\<}{\langle}
\renewcommand{\>}{\rangle}
\renewcommand{\vec}{\textbf}
\newcommand{\Li}{\mathcal L}

\newcommand{\Tr}{{\rm Tr\,}}

\graphicspath{ {images/} }

\newcommand{\tr}[1]{\operatorname{\textnormal{Tr}}\left( {#1} \right)} 
\newcommand{\trs}[2]{\operatorname{\textnormal{Tr}}_{{#1}}\left( {#2} \right)}  

\begin{document}

\title{Classical Models of Entanglement in Monitored Random Circuits}
\author{Oles Shtanko}
\author{Yaroslav A. Kharkov}
\author{Luis~Pedro Garc\'{i}a-Pintos}
\author{Alexey V. Gorshkov}
\affiliation{Joint Center for Quantum Information and Computer Science and Joint Quantum Institute, NIST/University of Maryland, College Park, Maryland 20742, USA}

\begin{abstract}
The evolution of entanglement entropy in quantum circuits composed of
Haar-random gates and projective measurements shows versatile behavior, with connections to phase transitions and complexity theory.
We 
reformulate the problem in terms of a classical Markov process for the dynamics of bipartition purities and establish a probabilistic cellular-automaton algorithm to compute entanglement entropy in monitored random circuits on arbitrary graphs.
In one dimension, we further relate the evolution of the entropy to a simple classical spin model that naturally generalizes a two-dimensional lattice percolation problem. We also establish a Markov model for the evolution of the zeroth R\'{e}nyi entropy and demonstrate that, in one dimension and in the limit of large local
dimension, it coincides with the corresponding second-R\'{e}nyi-entropy model.
 Finally, we 
 extend the Markovian description 
 to a more general setting that incorporates continuous-time dynamics, defined by stochastic Hamiltonians  and weak local measurements continuously monitoring the system.
\end{abstract}

\maketitle
 
Recent progress in creating and manipulating many-qubit devices
 led to their ability to arguably attain quantum supremacy, i.e.\ to exhibit dynamics that is not efficiently simulable on classical computers \cite{GoogleSupremacy,harrow2017quantum,Boixo2018}. Existing supremacy proposals include 
 circuit models that utilize randomly generated local gates to 
 create classically irreproducible quantum multi-qubit correlations \cite{bouland2018quantum,movassagh2019cayley}. Random circuits can be characterized by a rapid growth of entanglement across the system 
 \cite{dahlsten2007emergence,nahum2017quantum,KeyserlingkPRX2018,EmergentStatNahum2019}, and even though high entanglement alone does not guarantee hardness of classical simulation 
 \cite{terhal2002classical, AaronsonGottesman2004}, 
 the generation of entangled states is key to achieving such hardness in many cases
 \cite{vidal2004efficient}. Different behavior can be found if a random circuit is affected by measurements that project and thus disentangle individual qudits. Under persistent random measurements, a competition between entanglement production and reduction mechanisms leads to a phase transition~\cite{SzyniszewskiWeakMeas2019,EntanglTransYaodong2019,Zabalo2019,QuantZeno2018, SkinnerNahum2019,bao20,Cao2019SciPost,GullansHuse2019,gullans2019scalable,tang2020measurement,Choi2019,zhang2020nonuniversal,lopez2020mean} separating phases with area- and volume-law entanglement, similar to known dynamical quantum phase transitions in Hamiltonian systems \cite{dalessio2016quantum,vasseur2019entanglement,nandkishore2015MBL,abanin2019MBL}. Therefore, the evolution of entanglement in monitored random circuits exhibits versatile behavior associated with a range of fields from complexity theory to condensed matter physics.
 
 In this work, we 
 study the dynamics of entanglement 
in a wide class of random circuits under the combined effect of unitary gates and measurements. Starting with a random-walk model for entanglement growth in 1D unitary circuits, we generalize our approach to a 
Monte Carlo algorithm for computing the average entanglement entropy on arbitrary graphs in the presence of both unitary evolution and single-qudit measurements.
Furthermore, our approach can be mapped to a dynamical lattice percolation problem, which, in the limit of infinite qudit dimension, maps to a static percolation transition \cite{Zabalo2019, SkinnerNahum2019, GullansHuse2019}.
 We also study a continuous-time analogue of random circuits~\cite{BrownianCirc2019PRE,
SwinglePRX2019,
LongRange2019PRB,
zhou2019operator} in which the system evolves under a stochastic local Hamiltonian and is subject to dephasing and continuous weak measurements~\cite{wiseman_milburn_2009,jacobs_2014,Jacobs_2006}. 
We thus introduce the study of random quantum circuits to a rich arena of weak-measurement physics boasting both applied and foundational results
\cite{weber2014mapping,murch2013observing,
devoret2013superconducting,
PhysRevX.6.011002}.

Mapping of monitored quantum circuits to classical models has been studied recently using the replica trick for both closed \cite{EmergentStatNahum2019,napp2019efficient} and monitored \cite{Choi2019} systems.
The theoretical description is simplified 
in the limit of large qudit dimension $q \gg 1$ \cite{EmergentStatNahum2019, Choi2019}. 
In Refs.~\cite{nahum2017quantum,OperatorSpreadNahum2018}, it was shown that, in the large-$q$ limit, the evolution of R\'{e}nyi entropies  
in random unitary circuits can be described by a Kardar–Parisi–Zhang-type (KPZ-type) equation \cite{Kardar1986PRL}. 
Hydrodynamic descriptions of random unitary circuits   
are also discussed in Refs.~\cite{KeyserlingkPRX2018, Khemani2018PRX, Rakovszky2018PRX}.
In the presence of projective measurements, mapping to spin models connects the entanglement phase transitions to the percolation criticality and conformal field theories~\cite{Zabalo2019, CriticalityLudwig2020}.
The measurement-induced entanglement phase transition is also related to the encoder-decoder problem, with measurements playing the role of external noise corrupting the information \cite{GullansHuse2019, Choi2019}. Compared to previous approaches designed for 1D systems, our method
applies to arbitrary graphs, with significant speedups over exact simulation.

We consider a generic circuit acting on $q$-dimensional qudits and consisting of Haar-random gates and projective single-qudit measurements. As a measure of entanglement between a specified subsystem $A$ and the rest of the qudits, we use average R\'{e}nyi entropy  $S_2[A] \equiv -\<\log \Tr_A\bigl( \rho_A^2 \bigl)\>$, where $\<\dots\>$ denotes an average with respect to circuit realizations and $\rho_A$ is the reduced density matrix of $A$. 
In order to compute the average, we apply the \textit{annealed approximation},
$S_2[A] \simeq-\log P_A$, where $P_A \equiv \left\langle \Tr_A\bigl( \rho_A^2 \bigl) \right\rangle$ is the (average) purity. This approximation 
is exact for Haar-random states in the limit of large system size
\cite{nadal2010phase,nadal2011statistical,majumdar2014top}. We demonstrate in this work that this approximation is also remarkably accurate in application to random circuits.

To compute the R\'{e}nyi  entropy, we focus on 
the vector
$
\vec P =  \{P_{G}: G \in F\}
$
consisting of purities of all possible subsystems $G$ of the full system $F$. 
Below we mostly focus on the linear discrete-time Markovian evolution
\be\label{eq:Liouvillian_evolution}
\vec P(t+1) = \mathcal L\,\vec P(t),
\ee
where $\mathcal L$ is a time-dependent Liouvillian (see also Ref.\ \cite{kuo2019markovian}).  In the following sections, we will show the relevance of this model to random circuits. 

\textbf{Random Circuits.---}Consider the evolution of a pure state $|\psi\>$ under a single Haar-random gate $U$ supported on subset $\omega$, $|\psi'\> = U|\psi\>$.
The resulting (gate-averaged) purity of an arbitrary subset $G$ for the state $|\psi'\>$ is determined [see Supplemental Material (SM) \cite{supp}, Sec.\ I] by the purities for $|\psi\>$ via 
\be\label{eq:unitary_purity_evo}
P_G' = c_- P_{G\setminus \omega}+c_+P_{G\cup \omega},
\ee
where $c_- = d_1(d_2^2-1)/(d_\omega^2-1)$, $c_+ = d_2(d_1^2-1)/(d_\omega^2-1)$, $d_1 = d_{\omega\cap G}$, and $d_2 = d_{\omega\setminus G}$ \footnote{We refer to qudit sets using standard notation for the relative complement ($G\setminus G'$), union ($G\cup G'$), intersection ($G\cap G'$), as well as
Hilbert space dimension $d_G\equiv q^{|G|}$}.
For two-qudit gates with $d_1 = d_2 = q$, we obtain $c_-=c_+ = q/(q^{2}+1)$. Coefficients $c_+$ and $c_-$ are the only non-zero elements in the Liouvillian matrix $\mathcal L$ in Eq.~\eqref{eq:Liouvillian_evolution} representing a single random gate. For 
more than one gate, the total Liouvillian is the product of individual gate Liouvillians in reverse chronological order.

We now consider a projective measurement performed on a single qudit $\Omega$ of an observable 
$O = \sum_i o_i \Pi_i$, where $\Pi_i$ are projectors satisfying  
$\Pi_i\Pi_j=\Pi_i\delta_{ij}$. A measurement with outcome $o_i$ projects the system onto 
$|\psi'\> = \Pi_i|\psi\>/\sqrt{\<\psi|\Pi_i|\psi\>}$. Since measurements are always ``sandwiched" between Haar-random gates, 
without loss of generality, we replace the purities by their average 
over the basis of $O$, $P_G = \<\Tr\rho^2_G\>_O$. 
Applying the annealed approximation to $O_1/O_2$ 
$({\<O_1/O_2\>} \simeq {\<O_1\>/\<O_2\>}$), 
the post-measurement purity becomes 
\be\label{eq:measurement_evo}
\begin{split}
P_G' \simeq \frac{\left\<\mathcal \Tr_G(\Pi_i|\psi\>\<\psi|\Pi_i)_G^2\right\>_{O}}{\left\<\<\psi|\Pi_i|\psi\>^2\right\>_{O}} =\frac{P_{G\cup\Omega}+P_{G\setminus\Omega}}{1+P_{\Omega}}.
\end{split}
\ee

To express the combined non-linear evolution under Eqs.~(\ref{eq:unitary_purity_evo},\ref{eq:measurement_evo}) in terms of linear evolution, we define \textit{unnormalized purities} $\tilde{\vec{P}}$ evolving under $\tilde{\vec{P}}(t+1) = \mathcal L \tilde{\vec{P}}(t)$, where unitary dynamics is given by  Eq.~\eqref{eq:unitary_purity_evo}, while measurements are treated as the transformation $\tilde P'_G = \tilde P_{G\setminus\Omega}+\tilde P_{G\cup\Omega}$.
Then, after an arbitrary number of unitary and measurement transformations, the physical purity vector 
is
\be
P_G =\tilde P_G/\tilde P_F,
\ee
where $\tilde P_F$ is the unnormalized purity of the full system. 

\begin{figure*}[t!]
    \centering
    \includegraphics[width=1\textwidth]{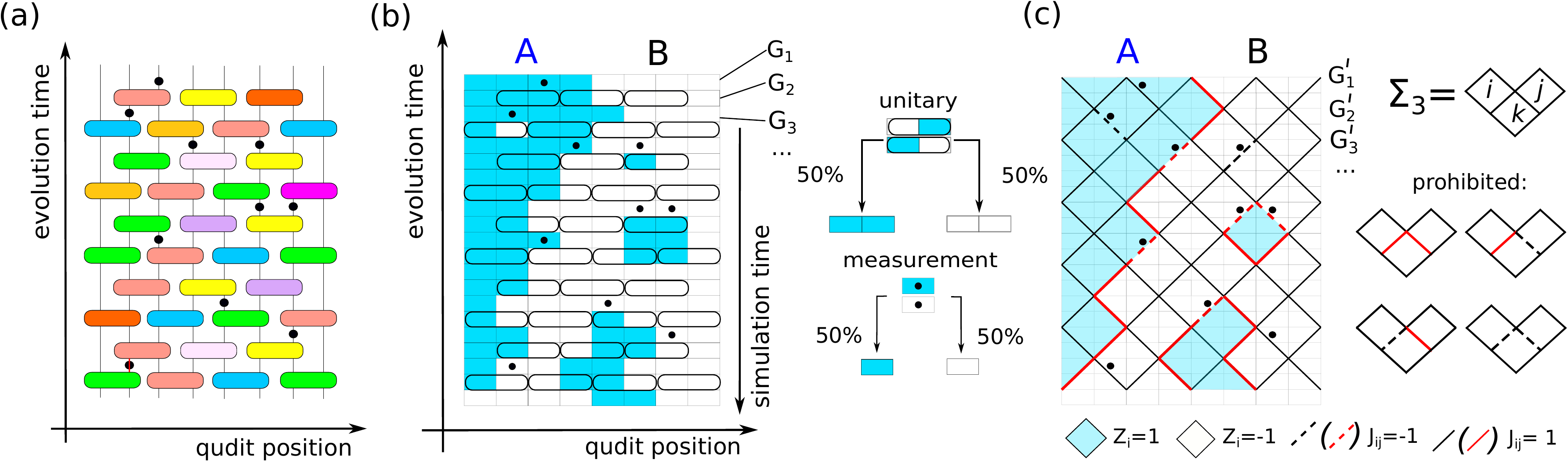}
    \caption{\textbf{1D brickwork circuits}. 
    (a) Monitored 1D brickwork circuit. 
    Rectangles represent two-qudit Haar-random unitaries, while black circles represent projective measurements. 
    (b) A trajectory generated by a probabilistic cellular automaton for the monitored 1D random circuit in (a). The process starts from the top (final time) and flows down following local rules described on the right for unitaries (ovals) and measurements (dots). The 20 simulation steps shown (from $G_1$ to $G_{21}$) correspond to the 10 brickwork layers interleaved with 10 measurement layers shown in (a). The resulting trajectory can be converted to a single contribution $\mathcal P^{MC}_{\mathcal G}$ to the sum in Eq.~\eqref{eq:Pdecomposition}. 
    (c) One configuration, corresponding precisely to the trajectory in (b), of a classical spin model whose partition function defines the final purity via Eq.~\eqref{eq:stat_model}. The Hamiltonian is the 2D Ising model in Eq.~\eqref{eq:stat_model_ham} for classical spins $Z_i$ associated with  squares ($Z_i = 1$ are blue, $Z_i =-1$ are white). Each edge 
    directly below a measurement has an additional degree of freedom $J_{ij}$: 
    $J_{ij} = -1$ (dashed) if the measurement changed the color of the cell in (b), and $J_{ij} =1$ (solid) otherwise. The operator $\mathcal P$ acts non-trivially only on the top row of cells,  projecting them onto the target bipartition $A$, and on all directed 3-sets $\Sigma_3$ prohibiting certain configurations of edges (shown on the right) irrespective of square colors.} 
    \label{fig:1dcircuit}
\end{figure*}

\begin{figure}[t!]
    \centering
    \includegraphics[width=0.5\textwidth]{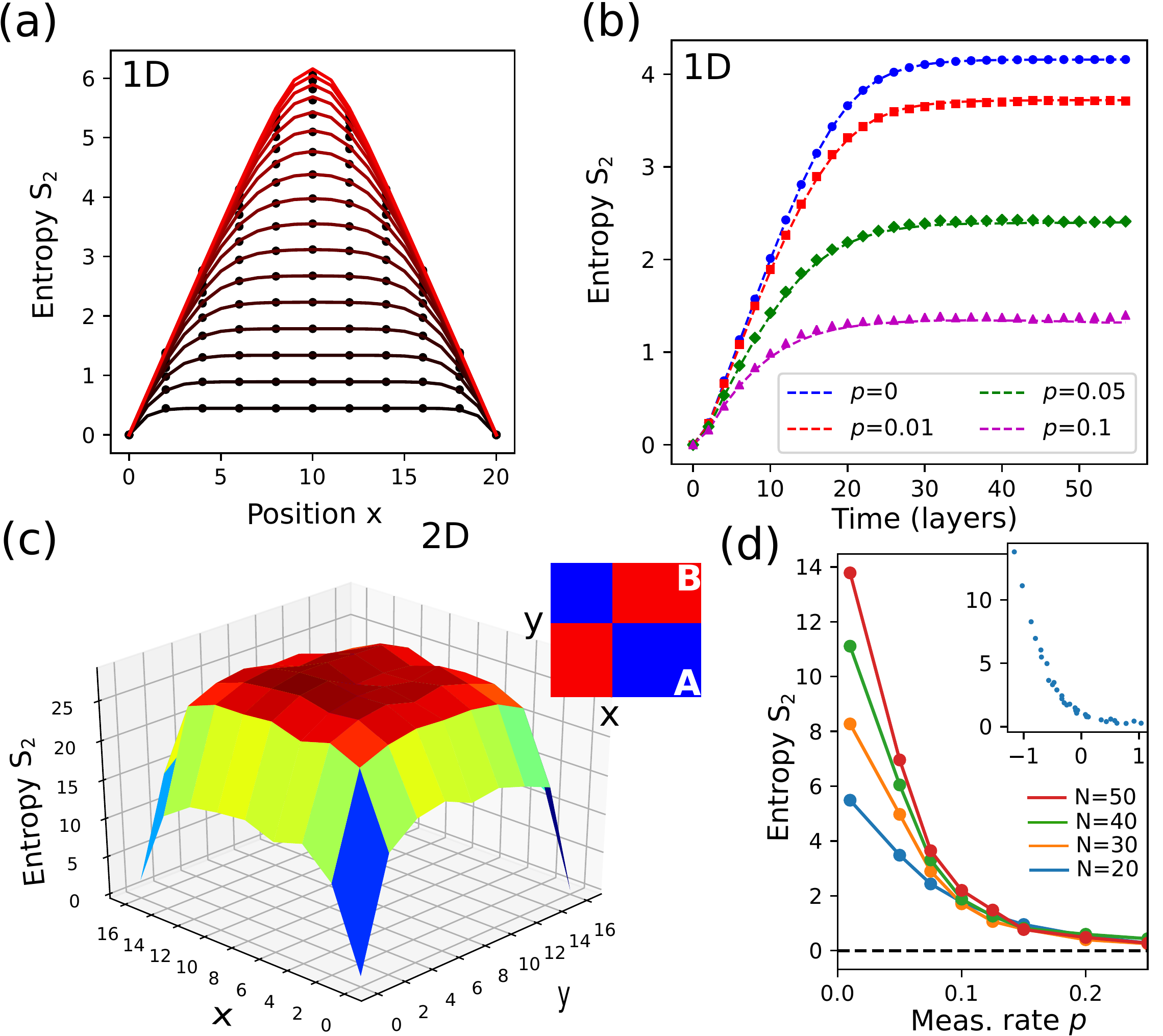}
    \caption{\textbf{Classical simulation ($q=2$)}. 
    (a) Comparison between the solution of 
    Eq.~\eqref{eq:kpz_equation} taken at  times $t=2n$, $n\in \mathbb Z$, (solid curves) and exact numerical simulation for a 1D circuit (dots) and even cut positions $x$, averaged over $10^2$ realizations; system size is $N=20$. 
    (b) Entropy growth in a 1D brickwork circuit with measurements. The figure compares exact numerics 
    averaged over $10^2$ gate-measurement realizations (dots) to a MCMC simulation using $10^4$ samples 
    for each of the 
    $10^2$ sampled measurement configurations (dashed curves) for system size $N=14$. 
    (c) Long-time ($t = 64$) value of the entropy for a monitored circuit acting on a 2D system of $16\times16$ qudits with the brickwork circuit arrangement introduced in Ref.~\cite{OperatorSpreadNahum2018} and with measurement probability $p=0.1$ for every qudit in every layer. 
    The entropy is shown as a function of system division (for cuts at even positions) illustrated in the inset and computed with the MCMC algorithm using $10^4$ samples 
    for each of the 
    150 sampled measurement configurations. 
    (d) Scaling of the entropy for a monitored 1D system as a function of the measurement rate $p$, as calculated using MCMC with up to $10^7$ samples. 
    The inset shows that all four curves collapse to a single curve when plotted as a function of
    $(p-p_c)N^{1/\nu}$ 
    for critical point $p_c \simeq 0.14(2)$  and critical exponent $\nu \simeq 1.7(2)$ 
    describing the area-law to volume-law transition.}
    \label{fig:plots}
\end{figure}

\textbf{Classical Models.---}The linearity of purity evolution enables the mapping of entanglement dynamics onto classical problems such as multi-particle random walks. As an illustration, consider a simple brickwork 1D unitary circuit [Fig.~\ref{fig:1dcircuit}(a) without measurements]. 
It follows from Eq.~\eqref{eq:unitary_purity_evo} that the dynamics of
bipartition purities $P^t_x$ parametrized by the cut position $x$
is a modified single-particle random-walk:
$
P^{t+1}_x = c\bigl(P^t_{x-1}+P^t_{x+1}\bigl)
$ for
$x = 2n+t({\rm mod}\;2)$, $n\in \mathbb Z$, and $c = q/(q^2+1)$. 
The
R\'{e}nyi entropies $S_2(t,x) = -\log P_x^t$ in the continuous limit then satisfy a noise-averaged KPZ equation
\be\label{eq:kpz_equation}
\partial_t S_2 = \mu\bigl(\partial^2_x S_2-(\partial_x S_2)^2\bigl)+\beta,
\ee
where $\beta=\log((q^2+1)/2q)$ and $\mu = \beta/(\log q)^2$ (see SM Sec.\ III \cite{supp}).  
The long-time asymptotic solution of this hydrodynamic equation, subject to vanishing boundary conditions at the endpoints of the 1D qudit chain, is
\be \label{eq:S2_asym}
S_2(t\to\infty,x)\to\log\left[\frac{\cosh{\left(\alpha N/2 \right)}}{\cosh{\left(\alpha(x- N/2)\right)}} \right],
\ee
where $\alpha =\sqrt{\beta/\mu} =  \log q$ and $N$ is the number of qudits. 
The maximum value of the entropy at the center of the chain,
$S_2(t\to\infty, x=N/2)\simeq \alpha N/2-\log 2+O(\exp(-\alpha N))$,
exhibits volume-law scaling. The precision of the hydrodynamic approximation for $q=2$ is illustrated in Fig.~\ref{fig:plots}(a). 

In more general settings, 
the number of partitions involved in the dynamics of purities $P_A$ (or $\tilde P_A$) 
typically grows exponentially in time.
To describe such processes, we consider a sequence of $n$ circuit layers, each described by a sparse Liouvillian $\mathcal L_k$, sorted in 
reverse chronological order. Let us define a collection of neighboring sets $\mathcal N_k[G] = \{G': \mathcal L^k_{G,G'}\neq 0\}$
and multiple trajectories $\mathcal G =\{G_1\to G_2\dots \to G_n\}$ such that $G_{k+1} \in \mathcal N_k[G_k]$, and $G_1 = A$. Assuming the system is initialized in a product state, 
i.e. $P_G(0) = 1$ for all $G$, the (unnormalized) purity after $n$ circuit layers is 
\be\label{eq:Pdecomposition}
\tilde P_{A} = \sum_{\mathcal G} P_{\mathcal G}, \qquad P_{\mathcal G} = \prod_{k=0}^{n-1} \Li^k_{G_k,G_{k+1}}.
\ee

The sum in Eq.~\eqref{eq:Pdecomposition} can be represented by a generalized random walk. Following a Markov chain Monte Carlo (MCMC) algorithm, instead of listing all possible trajectories, we consider their probabilistic generation, choosing $G_{k+1}$ 
from the set $\mathcal N_k[G_k]$ with probability $\pi^k_G = \Li^k_{G,G_k}/T_k$,
where $T_k =\sum_{G\in N_k[G_k]}\Li^k_{G,G_k} $.
In the MCMC scheme,
the purity $\tilde P_A$ in Eq. (\ref{eq:Pdecomposition}) is approximated by the sample average $\tilde P_A\approx \<P^{MC}_{\mathcal G}\>_{\mathcal G} = \< \prod_k T_k \>_\mathcal{G}$.
In practice, in order to estimate $\tilde P_A$, one needs an average over a large number of trajectories due to the broad distribution of $P^{MC}_\mathcal G$.
Nevertheless, simulations show that this number can be significantly smaller than the full number of trajectories $\mathcal G$.

For simulation purposes, the MCMC algorithm can be mapped to a probabilistic classical cellular automaton \cite{Fernandez2018}. 
It is instructive to start with the monitored 1D brickwork circuit shown in Fig.~\ref{fig:1dcircuit}(a). We describe each unitary and measurement layer by a Liouvillian $\Li^{\rm u}_k$ and $\Li^{\rm m}_k$, respectively.  
The colored pattern in Fig.~\ref{fig:1dcircuit}(b) illustrates the dynamics of the corresponding cellular automaton evolving from top to bottom generating a trajectory, where each cell represents a qudit.
Blue cells in each row denote the qudits forming $G_k\in \mathcal G$, where $\mathcal G$ is a trajectory formed by alternating $\Li^{\rm u}_k$ and $\Li^{\rm m}_k$.

Each unitary layer Liouvillian $\Li^{\rm u}_k$ can be converted into a rule:
all distinct-color pairs of cells affected by a Haar-random  two-qudit gate in this layer switch to the same color, chosen by a coin toss for each gate. 
Every such switch contributes a factor $T_k = 2q/(q^2+1)$ to the corresponding amplitude $P^{MC}_{\mathcal G}$.
A measurement layer $\Li^{\rm m}_k$ is equivalent to a color flip of each cell representing a measured qudit with probability $\pi^k_G = 1/2$, with no contribution to $P^{MC}_{\mathcal G}$, i.e.\ $T_k=1$. 
The method can be used as a fast sampling algorithm producing purities [see Fig.~\ref{fig:plots}(b)] and applies to arbitrary graphs 
[see Fig.~\ref{fig:plots}(c) showing MCMC results for 2D].
The MCMC method is particularly useful for analysing measurement-induced entanglement phase transitions. For example, Fig.~\ref{fig:plots}(b) illustrates entropy growth in a 1D brickwork circuit with measurements affecting each qudit with probability $p$ at every layer; MCMC shows excellent agreement with exact numerics. The algorithm can be used to reach larger system sizes enabling an estimate of the critical $p$ and critical exponents, as illustrated in Fig.~\ref{fig:plots}(d).

Instead of the MCMC algorithm, 
we can alternatively relate $\tilde P_A$ to a partition function of a classical Hamiltonian $H$, with trajectories $\mathcal G$ mapped to 
energy levels $E[\mathcal G] \in {\rm spec}(H)$,
\be\label{eq:stat_model}
\tilde P_A = \sum_\mathcal {G} \exp(-\beta E[\mathcal G])=\Tr\Bigl( \mathcal P \exp(-\beta H)\Bigl),
\ee
where $\beta= \log\bigl((q^2+1)/2q\bigl)$ is an effective inverse temperature, 
and $\mathcal P$ is a projector onto the levels $E[\mathcal G]$. 

For the monitored 1D brickwork circuit, such a classical model can be constructed as follows (see also SM Sec.\ IV \cite{supp}). 
Let us combine each pair of unitary and measurement layers into a single  Liouvillian  
$\Li^{\rm um}_k$ coupling $G'_k$ to $G'_{k+1}$, where the full trajectory $\mathcal G =\{G'_1\to G'_2\dots \to G'_n\}$ includes only odd-time bipartitions, $G_k' = G_{2k-1}$.  
Then, 
for any trajectory $\mathcal G$, bipartitions propagate on a square lattice shown in Fig.~\ref{fig:1dcircuit}(c),   
representing a classical analogue of the Feynman checkerboard \cite{feynman2010quantum}. 
We associate a classical variable $Z_i=1$ to a square if it is part of $G'_k$ [blue squares in Fig.~\ref{fig:1dcircuit}(c)], while the rest have $Z_i = -1$ (white squares); domain walls (red) separate blue squares from white ones.
Furthermore, any edge $(i,j)$ separating adjacent squares $i$ and $j$ and situated below a measurement in Fig.~\ref{fig:1dcircuit}(c)  
is assigned an additional degree of freedom $J_{ij}= 1$ (solid line) or $J_{ij} = -1$ (dashed line). 
For all other edges, we set $J_{ij}=1$.  
In this setting, the classical Hamiltonian and projector in Eq.~\eqref{eq:stat_model} are
\be\label{eq:stat_model_ham}
\begin{split}
&H = \frac 12 N_0-\frac 14\sum_{\<i,j\>}J_{ij}Z_iZ_j,\\
&\mathcal P = \mathcal P_A\prod_{i,j,k\in \Sigma_3}\Bigl(1-\frac 14(Z_k-J_{ik}Z_i)(Z_k-J_{jk}Z_j)\Bigl),
\end{split}
\ee
where $\<i,j\>$ is the sum over all $N_0$ edges, 
$\mathcal P_A = \prod_{i\in A} \frac 12(1+Z_i)\prod_{j\in F\setminus A}\frac 12(1-Z_j)$ 
projects the top squares onto the bipartition configuration $A$, and $\Sigma_3$ are all possible oriented 3-cell groups, as shown in Fig.~\ref{fig:1dcircuit}(c). This model applies to periodic boundary conditions (for open boundary conditions, see SM Sec.\ IV \cite{supp}).
The Hamiltonian $H$ is simply equal to the total number of solid red edges plus the total number of dashed black edges.

\textbf{Connection to Hartley entropy.---}The dynamics of the Hartley entropy, 
defined as the zeroth R\'{e}nyi entropy $S_0[A] = \log(R_A)$, where $R_G = {\rm rank}\,(\rho_G)$, is typically qualitatively different
from R\'{e}nyi entropies of order $n\geq1$ \cite{SkinnerNahum2019}. Nevertheless, we show that $S_0$ has deep connections to the 2nd R\'{e}nyi entropy $S_2$.
Under a generic unitary transformation supported on $\omega$, $G\cap\omega\neq \emptyset$ and $\omega\setminus G\neq \emptyset$, the rank follows non-liear Markovian dynamics
(see SM Sec.\ II \cite{supp}) 
\be \label{eq:S_0_unitary}
R'_G = \frac{R_{G\setminus\omega}R_{G\cup\omega}} {R_{G\Delta\omega}}\min(d_{G\cap\omega}R_{G\cap\omega},d_{\omega\setminus G}R_{\omega\setminus G}), 
\ee
where 
$G\Delta \omega \equiv (G\cup \omega)\setminus(G\cap \omega)$. The effect of a measurement on qudit $\Omega$ is
 \be\label{eq:S_0_meas}
R'_G = \frac{R_{G\setminus\Omega}R_{G\cup\Omega}} {R_{G\Delta\Omega}}\min(R_{G\cap\Omega},R_{\Omega\setminus G}).
\ee
In 1D systems, these expressions become linear: $R'_G = \min\bigl(d_{G\cap\omega}R_{G\setminus\omega},d_{\omega\setminus G}R_{G\cup \omega}\bigl)$ for unitary gates \cite{nahum2017quantum} and $R_G' = \min{(R_{G\setminus \Omega}, R_{G \cup \Omega})}$ for measurements. It follows from these expressions that finding $S_0$ for a  1D system initialized in a product state can be reduced to a minimization of $P_{\mathcal G}$ in Eq.~\eqref{eq:Pdecomposition}, $P_A = \min P_\mathcal G$. For 1D brickwork circuits, this is equivalent to path minimization
on a percolated lattice, as previously suggested 
in Ref.~\cite{SkinnerNahum2019}. Notably, 
$S_0$ can also be written as $
S_0[A] =\log(q) \min_{\mathcal G} E[{\mathcal G}]$, where $E[{\mathcal G}]$ is the spectrum of the Hamiltonian in Eq.~\eqref{eq:stat_model_ham} for configurations satisfying $\mathcal P[\mathcal G]=1$.
As a consequence, in the limit $q\to\infty$, entropies $S_0$ and $S_2$ coincide, given that, in the limit $\beta\to\infty$, the partition function in Eq.~\eqref{eq:stat_model} reduces to $S_2[A] \simeq \beta \min_{\mathcal G} E[\mathcal G]$ with $\beta\sim\log q$. 

In the hydrodynamic approximation, Hartley entropy for 1D unitary brickwork circuits evolves according to the continuum limit of Eq.\ (\ref{eq:S_0_unitary})
(see SM Sec.\ III \cite{supp}):
\be \label{eq:dt_G_0}
\partial_t S_0 = \frac{1}{2}\partial^2_{x}S_0 - |\partial_x S_0| + \alpha,
\ee
where $\alpha=\log q$. For product initial states, the time-dependent solutions of Eqs. (\ref{eq:dt_G_0}) and (\ref{eq:kpz_equation}) coincide in the limit $q\to\infty$ at scales $x\gg 1$. 
Furthermore, the stationary solution of Eq.~\eqref{eq:dt_G_0} for an initial product state and open boundary conditions [$S_0(x=0) = S_0(x=N) = 0$] is $S_0(t\to\infty, x) =  \alpha(N/2-|x-N/2|)$, exhibiting the same volume-law entanglement as $S_2$.

\textbf{Continuous Evolution.---}To complete the analysis of the evolution of purity, we propose a continuous-time version of monitored random circuits. 
We consider a local stochastic Hamiltonian $H(t) = \sum_i \mathbb I_{F\setminus \omega_i} \otimes h_i(t)$, where individual local terms $h_i(t)$ are stochastic Gaussian-unitary-ensemble matrices 
supported on respective sets $\omega_i$. The corresponding matrix elements $h_{i,\nu \mu}$ of $h_i$
satisfy $\< h^*_{i,\mu\nu}(t)h_{j,\mu'\nu'}(t')\>_h = \frac 12\alpha_i(t)d_{\omega_i}^{-1}\delta_{ij}\delta_{\mu\mu'}\delta_{\nu\nu'}\delta(t-t')$ for some positive functions $\alpha_i(t)$.
As a model for monitoring, we consider continuous weak measurements~\cite{PhysRevD.33.1643,diosi1988continuous,PhysRevB.63.115403,
Wiseman_1996} performed for arbitrary single-qudit physical observables $O_j$ (acting on site $\Omega_j$)
that yield a combined non-linear equation
\begin{align}
\label{eq:rho_monit_master}
\frac{d}{dt} \ket{\psi} =& -iH(t)|\psi\>\\
&+\sum_j \bigg[ - \kappa_j [\delta O_j(\psi)]^2 
+ \xi_j(t)\sqrt{2 \kappa_j} \delta O_j(\psi)\bigg] \ket{\psi},\nonumber
\end{align}
where $\delta O_j(\psi) = O_j - \bra{\psi}O_j\ket{\psi}$, $\xi_i(t)$ are independent stochastic variables satisfying $\<\xi_i(t)\xi_j(t')\>_\xi = \delta_{ij}\delta(t-t')$, and $\kappa_j$ characterize the strength of the coupling to the measurement apparatus~
\cite{wiseman_milburn_2009,jacobs_2014,Jacobs_2006}. 
Similar to
projective measurements, continuous measurements also reduce entanglement in the system, leading to a competition with unitary evolution~\cite{SzyniszewskiWeakMeas2019}. 
We note that what follows holds even if we replace our time-independent $O_j$ with $O_j(t) = V_j(t)O_jV^\dag_j(t)$, where $V_j(t)$ is an arbitrary time-dependent unitary, which changes the measurement basis.

We are looking for a closed set of equations for the purity vector $\vec P$, which evolves according to $\partial\vec P/\partial t = \mathcal{L}^{\rm u}_t(\vec P)+\mathcal{L}^{\rm m}_t(\vec P)$.
Here, the unitary Liouvillian obeys 
(see SM Sec.\ V \cite{supp})
\be\label{eq:cont_unit_evo}
 [\mathcal{L}^{\rm u}_t(\vec P)]_G = -\sum_i \alpha_i\Bigl(P_G+\frac {P_{G\Delta \omega_i}}{d_\omega}-\frac {P_{G\setminus\omega_i}}{d_{\omega_i\cap G}} -\frac {P_{G\cup\omega_i}}{d_{\omega_i\setminus G}} \Bigl).
\ee
As expected, for a single time-independent term $h_i$, the steady-state of the Liouvillian in Eq.~\eqref{eq:cont_unit_evo} yields the discrete-time evolution in Eq.~\eqref{eq:unitary_purity_evo} \cite{supp}.

The measurement Liouvillian obeys
\begin{align}\label{eq:cont_meas_main}
 &[\Li^{\rm m}_t(\vec P)]_G  = \sum_{j:\Omega_j\in  F\setminus G}\zeta_j C_{G,\Omega_j}+\sum_{j:\Omega_j\in  G}\zeta_j C_{F\setminus G,\Omega_j},  
\end{align}
where $\zeta_j = 8\kappa_j\Tr_{\Omega_j} O_j^2 / (q^2-1)$, 
the measure of correlation $C_{S,\Omega} = \left\langle \big\| \rho_{S\cup\Omega} - \rho_{S} \otimes \rho_{\Omega} \big\|_2^2 \right\rangle$ encodes the state of the system, and $\| \cdot \|_2$ is the 2-norm.
In the mean-field approximation, 
$
 C_{S,\Omega}  \simeq  P_{S\cup\Omega} - P_{S}P_{\Omega} 
$,
leading to the desired closed set of equations for purities. Importantly, in this approximation, when only one qudit is measured, the steady-state of the Liouvillian in  Eq.~\eqref{eq:cont_meas_main} yields the discrete-time annealed-approximation evolution in Eq.~\eqref{eq:measurement_evo} \cite{supp}.

\textbf{Outlook.---}In this work, we demonstrated that linear dynamics of purities arises in monitored random circuits and yields, under the annealed approximation, an
algorithm for computing entanglement based on a mapping to dynamical percolated lattices. 
Many classical spin models with local interactions have efficient classical solutions requiring only polynomial resources. Can we establish this property for dynamical percolated lattices explored in this work? If so, this work opens a pathway to establishing the approximate classical simulability of entanglement dynamics in
monitored random circuits. 
Also, having established in this work a close correspondence between zeroth and second R\'{e}nyi entropies, we conjecture such correspondence for other R\'{e}nyi entropies.

\textit{Note.---}During the preparation of this manuscript, a related preprint appeared 
\cite{fan2020self} studying 
purities to analyze entanglement phase transitions in 1D circuits using mappings to a quantum spin model.

\begin{acknowledgments}
\textbf{Acknowledgements}. We
thank Michael Gullans, Abhinav Deshpande, Pradeep Niroula, Zhicheng Yang, and Soonwon Choi for fruitful discussions. This work was supported by the NSF PFCQC program, DoE ASCR FAR-QC (award No.\ DE-SC0020312), DoE BES Materials and Chemical Sciences Research for Quantum Information Science program (award No.\ DE-SC0019449), DoE ASCR Quantum Testbed Pathfinder program (award No.\ DE-SC0019040), AFOSR, ARO MURI, ARL CDQI, and NSF PFC at JQI. 
\end{acknowledgments}

\bibliography{refs}


\clearpage

\pagebreak

\setcounter{page}{1}
\setcounter{equation}{0}
\setcounter{figure}{0}
\renewcommand{\theequation}{S.\arabic{equation}}
\renewcommand{\thefigure}{S\arabic{figure}}
\renewcommand*{\thepage}{S\arabic{page}}

\onecolumngrid

\begin{center}
{\large \textbf{Supplemental Material for \\``Classical Models of Entanglement in Monitored Random Circuits"}}\\
\vspace{0.25cm}
{Oles Shtanko, Yaroslav A. Kharkov, Luis~Pedro Garc\'{i}a-Pintos, and Alexey V. Gorshkov}
\end{center}

In this Supplemental Material, we present details omitted in the main text. In particular, in Sec.\ I, we derive Eqs.~(\ref{eq:unitary_purity_evo},\ref{eq:measurement_evo}) in the main text and analyze the accuracy of the annealed approximation. In Sec.\ II, we derive Eqs.~(\ref{eq:S_0_unitary},\ref{eq:S_0_meas}) in the main text. In Sec.\ III, we derive Eqs.\ (\ref{eq:kpz_equation},\ref{eq:S2_asym},\ref{eq:dt_G_0}) in the main text. In Sec.\ IV, we derive Eqs.~(\ref{eq:stat_model},\ref{eq:stat_model_ham}) in the main text. Finally, in Sec.\ V, we derive Eqs.~(\ref{eq:cont_unit_evo},\ref{eq:cont_meas_main}) in the main text, analyze the accuracy of the mean-field approximation used to simplify Eq.\ (\ref{eq:cont_meas_main}), and show that Eqs.~(\ref{eq:cont_unit_evo},\ref{eq:cont_meas_main}) can be used to derive Eqs.~(\ref{eq:unitary_purity_evo},\ref{eq:measurement_evo}).   

\section{Section I: Evolution of R\'{e}nyi entropies}

In this section, we derive the evolution Equations~(\ref{eq:unitary_purity_evo},\ref{eq:measurement_evo}) for averaged subsystem purities. We also analyze the accuracy of the annealed approximation used in the main text to connect R\'{e}nyi entropies and averaged purities.

\begin{figure}[t!]
    \centering
    \includegraphics[width=1\textwidth]{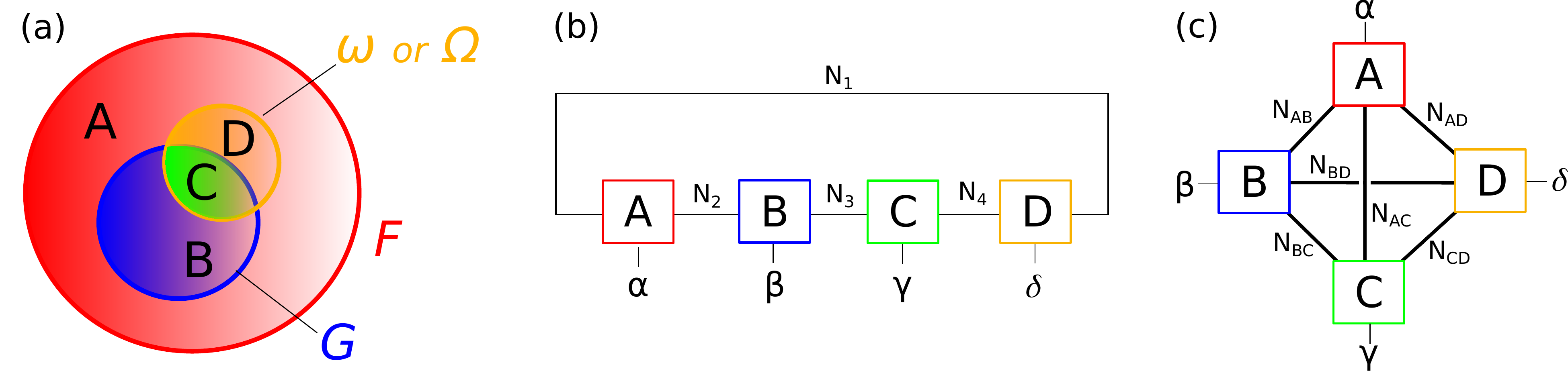}
    \caption{\textbf{System subdivision and matrix-product-state representation}. 
   (a) Subdivision of the full system $F$ into four 
    subsets $A = F\setminus(G\cup\omega)$, $B = G\setminus\omega$, $C =\omega\cap G$, and $D = \omega\setminus G$. 
    (b) Matrix product structure in 1D systems corresponding to Eq.~\eqref{eqs:mps1d}. Each square represents a matrix, while $N_i$ are local bond dimensions. 
    (c) Most general MPS structure, where $N_{GG'}$ represents the local bond dimension between sets $G$ and $G'$.}
    \label{figS1}
\end{figure}

\textbf{Unitary gates}. 
 Consider a single unitary gate $U$ supported on subset $\omega$. We are interested in finding the effect of $U$ on averaged purities. To perform this calculation, we divide the system into four subsets $A$, $B$, $C$, and $D$, as shown in Fig.~\ref{figS1}(a), and introduce the corresponding matrix representation for the density operator $\rho = \sum_{\rm ind} \rho_{\alpha \beta \gamma\delta,\alpha' \beta'\gamma'\delta'}|\alpha\beta\gamma\delta\>\<\alpha'\beta'\gamma'\delta'|$ and for the unitary gate $U = \sum_{\gamma\delta,\gamma'\delta'}U_{\gamma\delta\,\gamma'\delta'}|\gamma\delta\>\<\gamma'\delta'|$. Here labels $\{\alpha,\beta,\gamma,\delta\}$ refer to the basis states of $\{A,B,C,D\}$, respectively. Then 
\be
\begin{split}
P_G' &= \Bigl\<\Tr_{G}\bigl(\Tr_{F\setminus G} (U\rho U^\dag) \bigl)^2\Bigl\>_{{\rm circ},U}\\
&=\sum_{{\alpha_k,\beta_k,\gamma_k,\delta_k}} \Bigl\<\rho_{\alpha_1 \beta_1 \gamma_2\delta_2,\alpha_1\beta_2\gamma_3\delta_3}\rho_{\alpha_2 \beta_2 \gamma_5\delta_5;\alpha_2\beta_1\gamma_6\delta_6}\Bigl\>_{\rm circ}\Bigl\<U_{\gamma_1 \delta_1,\gamma_2\delta_2}U^*_{\gamma_4\delta_1,\gamma_3\delta_3}U_{\gamma_4\delta_4,\gamma_5\delta_5}U^*_{\gamma_1 \delta_4,\gamma_6\delta_6}\Bigl\>_{U},
\end{split}
\ee
where $\<\dots\>_U$ is the Haar average over $U$, while the average over the circuit $\<\dots\>_{\rm circ}$ includes the average over circuit evolution preceding the application of $U$. For random circuits, the $\<\dots\>_U$ and $\<\dots\>_{\rm circ}$ averaging can be separated because gate unitaries are sampled independently.

Correlation functions of Haar-random unitary matrix elements satisfy
\be\label{eqs:haar_corr_func}
   \langle U_{a,b} U^*_{c,d} U_{a',b'} U^*_{c',d'} \rangle_U = \frac{1}{d_{\omega}^2-1} \Bigl( \delta_{ac}\delta_{a'c'}\delta_{bd}\delta_{b'd'} + 
   \delta_{ac'}\delta_{a'c}\delta_{bd'}\delta_{b'd} - \frac{1}{d_\omega}(\delta_{ac}\delta_{a'c'}\delta_{bd'}\delta_{b'd} +
   \delta_{ac'}\delta_{a'c}\delta_{bd}\delta_{b'd'} ) \Bigl).
\ee
Denoting $d_1 \equiv d_{G\cap\omega}$ and $d_2 \equiv d_{\omega\setminus G}$, satisfying $d_\omega = d_1d_2$,
we obtain  
\be
P'_G=\frac{1}{d_\omega^2-1}\Bigl(d_1d_2^2 P_{G\setminus \omega}+d_2d_1^2 P_{G\cap \omega}-\frac 1{d_\omega}\Bigl(d_2d^2_1 P_{G\setminus \omega}+d_1d^2_2 P_{G\cap \omega}\Bigl)\Bigl)=\frac{1}{d_\omega^2-1}\Bigl(d_1(d_2^2-1)P_{G\setminus \omega}+d_2(d_1^2-1)P_{G\cup \omega}\Bigl),
\ee
which is Eq.~\eqref{eq:unitary_purity_evo} in the main text. This expression works for both pure and mixed states of the full system.

\textbf{Measurements}. Consider a projective measurement of qudit $\Omega$. We express this measurement using a projector $\Pi = U \Pi_0 U^\dag$, where $\Pi_0^2=\Pi_0$, $\Tr \Pi_0 = 1$, and $U$ is a Haar-random unitary supported on $\Omega$. We assume that the full system is in a pure state, $\rho = |\psi\>\<\psi|$. 
Using the annealed approximation to take the expectation value of a ratio, we obtain
\be\label{eqs:annld_frac}
P_G' \approx \frac{\left\<\mathcal \Tr_G(\Pi|\psi\>\<\psi|\Pi)_G^2\right\>_{\Pi}}{\left\<\<\psi|\Pi|\psi\>^2\right\>_{\Pi}} = \frac{\tilde P'_G}{\tilde P'_F}.
\ee
Let us assume first that $\Omega\in G$. Then
\be\label{eqs:purity_evo_derivation}
\begin{split}
\tilde P_G' &\equiv\Bigl\<\Tr_G(\rho_G \Pi)^2\Bigl\>_{\rho,\Pi} = \Bigl\<\Tr\Bigl(\rho U \Pi_0 U^\dag \rho U \Pi_0 U^\dag\Bigl)\Bigl\>_{\rho,U} = \\
& = \frac{1}{d_\Omega^2-1}\Bigl\<\Bigl(\Tr_G\rho_G^2 (\Tr_\Omega\Pi_0)^2+\Tr_{G\setminus \Omega}(\Tr_\Omega\rho_G)^2 \Tr_\Omega\Pi_0^2-\frac 1{d_\Omega}\Bigl(\Tr_{G\setminus \Omega}(\Tr_\Omega\rho)^2\Tr_\Omega\Pi_0^2+(\Tr_G\rho)^2(\Tr_\Omega\Pi_0)^2\Bigl)\Bigl) \Bigl\>_\rho\\
&= \frac{1}{d_\Omega+1}\Bigl(P_G+P_{G\setminus\Omega}\Bigl),
\end{split}
\ee
where we used the correlation function in Eq.~\eqref{eqs:haar_corr_func}.

The case when $\Omega\notin G$ can be approached in a similar fashion. 
Making use of the symmetry $\tilde P'_{F\setminus G} = \tilde P_G'$ for pure states, it is convenient to consider the purity $\tilde P'_{F\setminus G}$ of the complement of $G$ instead of $\tilde P'_G$ itself. Since $\Omega\in F\setminus G$, we can then use Eq.~\eqref{eqs:purity_evo_derivation} to obtain
\be
\tilde P'_G = \tilde P'_{F\setminus G}= \frac{1}{d_\Omega+1}\Bigl(P_{F\setminus G}+P_{(F\setminus G)\setminus\Omega}\Bigl) = \frac{1}{d_\Omega+1}\Bigl(P_G+P_{G\cup\Omega}\Bigl).
\ee
Both cases can be summarized as
\be\label{eqs:meas_puritie_num}
\tilde P_G'= \frac{1}{d_\Omega+1}\Bigl(P_{G\setminus\Omega}+P_{G\cup\Omega}\Bigl).
\ee
The denominator of Eq.~(\ref{eqs:annld_frac}) can also be derived from Eq.~\eqref{eqs:purity_evo_derivation} by taking $G=F$:
\be\label{eqs:meas_puritie_denom}
\begin{split}
\tilde P_F' = \frac{1}{d_\Omega+1}\Bigl(P_F+P_{F\setminus\Omega}\Bigl) = \frac{1}{d_\Omega+1}\Bigl(1+P_\Omega\Bigl).
\end{split}
\ee
Inserting Eqs.~\eqref{eqs:meas_puritie_num} and \eqref{eqs:meas_puritie_denom} into Eq.~\eqref{eqs:annld_frac}, we obtain Eq.~\eqref{eq:measurement_evo} in the main text. 

\begin{figure}[t!]
    \includegraphics[width=1\textwidth]{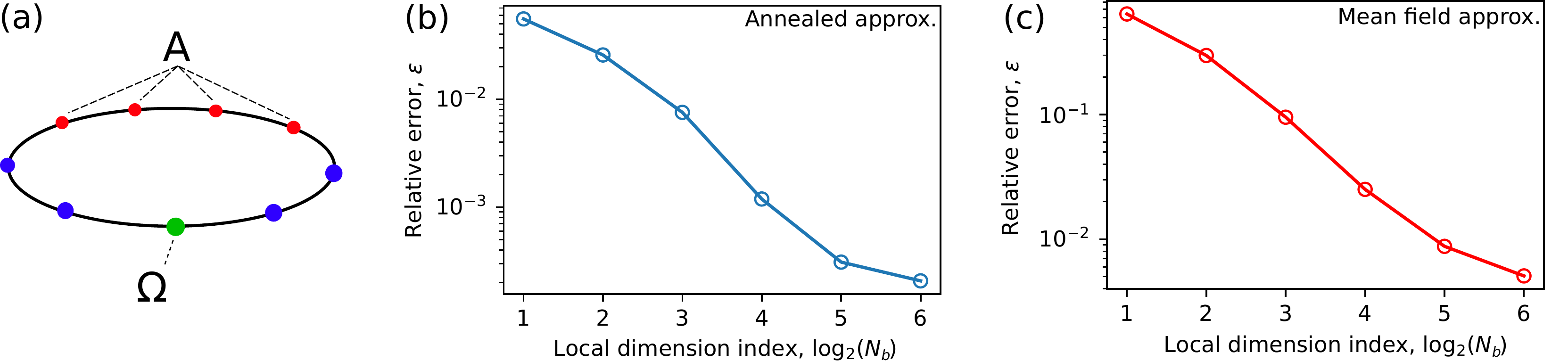}
\caption{ \textbf{Numerical analysis of approximations}. 
(a) The system consists of 9 qubits forming a 1D chain with periodic boundary conditions. The system is divided into a 4-qubit subsystem of interest $A$, the measured qubit $\Omega$, and the rest of the system. 
(b) Relative error of the annealed  approximation in Eq.~\eqref{eqs:annealed_approx} for the R\'{e}nyi entropy. The error is quantified as $\epsilon = \bigl|S_2[A]-S_2'[A]\bigl|/{S_2[A]}$, where $S_2[A] = -\<\log(\Tr_A \rho_A^2)\>$ and $S'_2[A] =-\log(\<\Tr_A \rho_A^2\>) $. The curve shows the error as a function of of the logarithm $\log_2 (N_b)$ of the local bond dimension $N_b$.
(c) Relative error of the mean-field approximation for the correlation distance measure in Eq.~\eqref{eqs:mf_corr_dis}. The error is quantified as $\epsilon = |C^1_{A,\Omega}-C^2_{A,\Omega}|/C^1_{A,\Omega}$, where $C^1_{A,\Omega} = \left\langle \big\| \rho_{A\cup\Omega} - \rho_{A} \otimes \rho_{\Omega} \big\|_2^2 \right\rangle$ and $C^2_{A,\Omega}= P_{A\cup\Omega}-P_A P_\Omega $.  }
\label{figS2}
\end{figure}

 As mentioned in the main text, the approximation
\be\label{eqs:annealed_approx}
S_2[A] = -\bigl\<\log( \Tr_A\rho_A^2)\bigl\> \simeq -\log(\bigl\< \Tr_A\rho_A^2\bigl\>) = -\log P_A
\ee
gives a good estimate of the average entropy $S_2(A)$.
The accuracy of this approximation is illustrated in Fig.~\ref{figS2}(b), using as a model a 1D system in a random matrix product state,
$
|\psi\> = \sum_{s_1\dots s_k}\Tr(A^{s_1}_1\dots A^{s_k}_k)|s_1\dots s_k\>,
$
where $A^{s_i}_i$ are $N_b\times N_b$ random matrices sampled from the Gaussian unitary ensemble
with local bond dimension $N_b$ [$x$ axis in Fig.~\ref{figS2}(b)]. Fig.~\ref{figS2}(b) suggests that the annealed approximation remains accurate even for states with small bond dimension, producing an error as low as $\epsilon\sim 10^{-2}$ for $N_b\sim 1$.

\section{Section II: Evolution of Hartley Entropies}

In this section, we derive the evolution Equations (\ref{eq:S_0_unitary},\ref{eq:S_0_meas}) for subsystem rank (whose logarithm is the Hartley entropy). 

To study the evolution of the rank, it is convenient to use a tensor network representation of the system state. We divide the system into four subsets shown in Fig.~\ref{figS1}(a) and associate with each subset a corresponding identically named tensor $A^\alpha$, $B^\beta$, $C^\gamma$, or $D^\delta$, where indices $\alpha$, $\beta$, $\gamma$, and $\delta$ run over corresponding basis states of each subset of qudits.
The structure of the tensors will depend on the dimensionality of the system and other properties of the circuit structure.

Let us start with a simple calculation for a 1D setting. For 1D systems, the tensor network is a matrix product state (MPS) representation of the system that can be written as
\be\label{eqs:mps1d}
\begin{split}
|\psi\> = \sum_{n_1\dots n_4,\alpha\beta\gamma\delta} A_{n_1n_2}^\alpha B_{n_2n_3}^\beta C_{n_3n_4}^\gamma D_{n_4n_1}^\delta |\alpha\beta\gamma\delta\>,
\end{split}
\ee
where indices $n_i=1,\dots, N_i$
run over the $N_i$ basis states of the bond connecting two of the subsystems, as shown in Fig.~\ref{figS1}(b).
Now consider a transformation $V$ supported on subsystem $\omega$ composed of 
$C$ and $D$. The MPS representation of the resulting state can be written as
\be
\begin{split}
V|\psi\> = \sum_{\vec n,\alpha\beta\gamma\delta} A_{n_1n_2}^\alpha B_{n_2n_3}^\beta \tilde C_{n_3n'_4}^\gamma \tilde D_{n'_4n_1}^\delta 
|\alpha\beta\gamma\delta\>,
\end{split}
\ee
where $n_4'=1,\dots,N'_4$ is expressed through the new bond dimension $N_4'$ between tensors $\tilde C$ and , and the new MPS matrices $\tilde C$ and $\tilde D$ are connected to the initial ones by the relation
\be
\sum_{n_4'}\tilde C_{n_3n_4'}^\gamma \tilde D_{n'_4n_1}^\delta = \sum_{n_4,\gamma'\delta'}V_{\gamma\delta,\gamma'\delta'}C_{n_3n_4}^{\gamma'} D_{n_4n_1}^{\delta'}.
\ee
If $V$ is a unitary operator, this relation contains $n_s = N_1N_3d_{G\cap\omega}d_{\omega\setminus G}$ equations. The matrix $\tilde C$ has $d_{G\cap\omega}N_{3}N'_{4}$ degrees of freedom (i.e\ indices), while the matrix $\tilde D$ has $d_{\omega\setminus G}N_{1}N'_{4}$ degrees of freedom. Importantly, $N_4^{'2}$ of these indices are redundant due to the existence of a gauge transformation $\tilde C' = \tilde C M$ and $\tilde D' = M^{-1}\tilde D $ for any invertible $N_4'\times N_4'$ matrix $M$. As a result, for the number of equations to coincide with the number of independent variables, the new rank $N_4'$ must satisfy
\be
d_{G\cap\omega}d_{\omega\setminus G}N_{1}N_{3} - d_{G\cap\omega}N_{1}N'_{4}-d_{\omega\setminus G}N_{3}N'_{4}+{N_4'}^2 = 0.
\ee
This expression can be rewritten as
$
(N_4'-d_{G\cap\omega}N_{3})(N_4' -d_{\omega\setminus G}N_{1})=0,
$ an equation with two positive roots. One of the solutions can be used to compute the rank of the subset $G$,
\be
R'_{G} ={\rm rank}(B^\beta \tilde C^\gamma) = N_2N_4'.
\ee
Taking into account that $R_{G\setminus \omega} = {\rm rank}(B^\beta) =  N_2N_3$ and $R_{G\cup\omega} = {\rm rank}(B^\beta C^\gamma D^\delta) = N_1N_2$, we obtain
\be\label{eqs:S0_unitary_formula}
R_G' = \min\bigl(d_{G\cap\omega}R_{G\setminus\omega},d_{\omega\setminus G}R_{G\cup \omega}\bigl),
\ee
which is Eq.~\eqref{eq:S_0_unitary} in the main text.
In the case when $d_{G\cap\omega} = d_{\omega\setminus G} =q$ the formula simplifies to
\be
R_G' = q\min\bigl(R_{G\setminus\omega},R_{G\cup \omega}\bigl),
\ee
leading to the following expression for the Hartley entropy:
\be
S_0[G] = \min\bigl(S_0[{G\setminus\omega}],S_0[{G\cup \omega}]\bigl)+\log q.
\ee

Consider now a measurement that projects the system onto state $|\psi'\> = V|\psi\>/\sqrt{\<\psi|V|\psi\>}$, where $V$ is supported on subset $\Omega$. We consider the same subdivision of the system up to the replacement $\omega\to\Omega$. Without loss of generality, the transformation $V$ can be chosen to be a projector on a subset of $\Omega$ of the form $V = I_{AB}\otimes |00\>\<00|_{CD}$ . Therefore, $\tilde B^\beta=0$ and $\tilde C^\gamma = 0$ for any $\beta,\gamma\neq 0$. The zeroth components of these tensors are connected to the initial MPS representation via
\be \label{eq:zeroth}
\sum_{n_4'}\tilde C_{n_3n_4'}^0 \tilde D_{n'_4n_1}^0 = {\rm const}\times \sum_{n_4}C_{n_3n_4}^{0} D_{n_4n_1}^{0},
\ee
where the constant arises due to normalization.

Using analysis similar to the case of unitary $V$, we find that Eq.\ (\ref{eq:zeroth}) contains $N_{1}N_{3}$ equations for $N_{3}N'_{4}+N_{1}N'_{4}-{N_4'}^2$ variables. Hence, the new rank satisfies 
\be\label{eqs:S0_meas_formula}
R_G' = \min\bigl(R_{G\setminus\Omega},R_{G\cup \Omega}\bigl),
\ee
which is Eq.~\eqref{eq:S_0_meas} in the main text.

These results can be generalized to the more general (beyond 1D) tensor structure shown in Fig.~\ref{figS1}(c). 
The MPS then contains four-index tensors
capturing the most general structure of the state:
\be
\begin{split}
|\psi\> = \sum_{\vec n,\alpha\beta\gamma\delta} A_{n_{AB},n_{AC},n_{AD}}^\alpha B_{n_{AB},n_{BC},n_{BD}}^\beta C_{n_{AC},n_{BC},n_{CD}}^\gamma D_{n_{AD},n_{BD},n_{CD}}^\delta |\alpha\beta\gamma\delta\>,
\end{split}
\ee
where indices $n_{GG'}=1, \dots, N_{GG'}$ run through the $N_{GG'}$ basis state of the bond between sets $G$ and $G'$.

Similarly to the 1D case, we consider a transformation $V$ that affects subsystem $\omega$ and changes tensors $C$ and $D$. The new tensors are related to the original ones via
\be \label{eq:tensors4}
\sum_{n_{CD}'}\tilde C_{n_{AC},n_{BC},n'_{CD}}^\gamma \tilde D_{n_{AD},n_{BD},r'_{CD}}^\delta = \sum_{n_{CD},\beta'\gamma'}V_{\gamma\delta,\gamma'\delta'}C_{n_{AC},n_{BC},n_{CD}}^{\gamma'} D_{n_{AD},n_{BD},n_{CD}}^{\delta'}.
\ee
Analysis similar to the 1D case applies here. Equation (\ref{eq:tensors4} contains $n_s = d_{G\cap\omega}d_{\omega\setminus G} N_{AC}N_{BC}N_{AD}N_{BD}$ equations for the $d_{G\cap\omega}N_{AC}N_{BC}N'_{CD}$ variables in tensor $\tilde C$ and the  $d_{\omega\setminus G}N_{AD}N_{BD}N_{CD}'$ variables in tensor $\tilde D$. Excluding ${N^{'2}_{CD}}$ redundant variables, we find that $N_{CD}'$ satisfies
\be
N_{CD}^{'2}-d_{G\cap\omega}N_{AC}N_{BC}N'_{CD}-d_{\omega\setminus G}N_{AD}N_{BD}N'_{CD}+d_{G\cap\omega}d_{\omega\setminus G} N_{AC}N_{BC}N_{AD}N_{BD}=0,
\ee
which can be rewritten as
\be \label{eq:NpCD}
(N_{CD}' - d_{G\cap\omega}N_{AC}N_{BC})(N_{CD}'-d_{\omega\setminus G}N_{AD}N_{BD})=0,
\ee
which has two positive roots. Taking into account that
\be
R'_G = {\rm rank}(B^\beta C^\gamma) = N_{AB}N_{BD}N_{AC}N'_{CD}
\ee
and  
\be
R_{G\setminus\omega} = {\rm rank}(B^\beta) = N_{AB}N_{BC}N_{BD},\quad R_{G\cup \omega} = {\rm rank}(B^\beta C^\gamma D^\delta) = {\rm rank}(A^\alpha) = N_{AB}N_{AC}N_{AD},
\ee
and taking the smaller root in Eq.\ (\ref{eq:NpCD}), we obtain the following equation for the rank:
\be\label{eqs:RG_imtermediate}
R'_G = \min (d_{G\cap\omega}N^2_{AC}R_{G\setminus\omega},d_{\omega\setminus G}N_{BD}^2R_{G\cup\omega}).
\ee
To express quantities such as $N^2_{AC}$ and $N_{BD}^2$, one needs expressions for the rank corresponding to other subsets of the system:
\be
\begin{split}
 &R_{ G\cap\omega} = {\rm rank}(C^\gamma) = N_{AC}N_{BC}N_{CD},\\
 &R_{\omega\setminus G} = {\rm rank}(D^\delta) = N_{AD}N_{BD}N_{CD},\\
 &R_{G\Delta\omega} = {\rm rank}(A^\alpha C^\gamma) = {\rm rank}(B^\beta D^\delta) = N_{AB}N_{AD}N_{BC}N_{CD}.
 \end{split}
\ee
It follows from these expressions that
\be
N_{BD}^2 = \frac{R_{G\setminus\omega}R_{\omega\setminus G}}{R_{G\Delta\omega}}, \qquad N_{AC}^2 = \frac{R_{\omega\setminus G}R_{ G\cap\omega}}{R_{G\Delta\omega}}.
\ee
Substituting these equalities into Eq.~\eqref{eqs:RG_imtermediate}, we obtain the final expression connecting the rank of subspace $G$ after the application the unitary $V$ to ranks before the application:
\be
R_G' = \frac{R_{G\setminus\omega}R_{G\cup\omega}}{R_{G\Delta\omega}}\min(d_{G\cap\omega}R_{G\cap\omega},d_{\omega\setminus G}R_{\omega\setminus G}).
\ee
A similar expression can be obtained for the effect of a measurement:
\be
R_G' = \frac{R_{G\setminus\Omega}R_{G\cup\Omega}}{R_{G\Delta\Omega}}\min(R_{G\cap\Omega},R_{\Omega\setminus G}),
\ee
where, as before, $\Omega$ is the measured set.

In 1D systems, the evolution described by Eqs.~\eqref{eqs:S0_unitary_formula} and~\eqref{eqs:S0_meas_formula} can be mapped to the minimum cut problem on a square lattice. To show this, we use the trajectory representation we defined for purities, but now formulated for ranks $R_G$. We formally write the evolution in the form
\be\label{eqs:r_liouv}
R'_{G} = \min_{G}(\mathcal L_{G,G'} R_{G'}),
\ee
where the Liouvillian $\mathcal L$ has the same structure as for the evolution of physical purities $P$ given by Eq.~\eqref{eq:unitary_purity_evo} and the numerator of Eq.~\eqref{eq:measurement_evo},
with $c_-=d_{G\cap\omega}$ and $c_+ = d_{\omega\setminus G}$.

As for purities, we consider all possible trajectories $\mathcal G =\{G_1\to G_2\dots \to G_n\}$ such that $G_{k+1} \in \mathcal N_k[G_k]$, and $G_1 = A$. Then, if the system is initialized in a product state, $R_G(0)=1$, the evolution can be presented in the form
\be\label{eqs:r_product}
R_A = \min_{\mathcal G}R_{\mathcal G}, \qquad R_{\mathcal G} = \prod_{k=1}^{n} \mathcal L^k_{G_k,G_{k+1}}.
\ee
The proof of this expression can be carried out by induction. First, the expression in Eq.~\eqref{eqs:r_liouv} is equivalent to Eq.~\eqref{eqs:r_product} under the condition that initially $R_{G} = 1$ for any $G$. Next, we assume that Eq.~\eqref{eqs:r_product} holds for a circuit of depth $n-1\geq 2$ described by Liouvillians $\{\mathcal L_2,\dots \mathcal L_{n}\}$ and consider different $n$-step trajectories $\mathcal G[G']$ starting from the set $G'$. Then
\be
R^{n+1}_A = \min_{G'}\Bigl(\mathcal L^{1}_{G,G'} \prod_{G_k\in \mathcal G[G']}^{n} \mathcal L^k_{G_k,G_{k+1}}\Bigl) = \min_{\mathcal G}R_{\mathcal G}.
\ee
In the case of a 1D brickwork configuration, the rank can be expressed using the same statistical model as in Eq.~\eqref{eq:stat_model_ham}, 
\be\label{eqs:stat_model}
R_A = \min_{\mathcal G} \mathcal \exp(-\alpha E[\mathcal G]),
\ee
where $\alpha=\log q$ and $E[\mathcal G]$ are the energies of all allowed configurations corresponding to $\mathcal G$ [see Eqs.~\eqref{eqs:unitary_mapping}-\eqref{eqs:two_meas_mapping}]. The minimum configuration energy for the statistical model is then given by the minimum total length of domain walls excluding all dashed segments on a percolated square lattice (see Sec.\ IV for more details). In the case of a bipartition, this reduces to a minimum path configuration connecting the cut at the top of the lattice to its other boundaries.

\section{SECTION III: Hydrodynamic equations for entropies in 1D}

In this section, we derive the 1D KPZ equation for R\'{e}nyi entropies [Eq.~\eqref{eq:kpz_equation} in the main text] and its analytical solution [Eq.~\eqref{eq:S2_asym}], as well as the continuous evolution equation for Hartley entropies [Eq.~\eqref{eq:dt_G_0}]. 

Let us start from the formal solution of Eq.~\eqref{eq:kpz_equation}. Substituting the definition of the entropy $S_2(t,x)=-\log P_c(t,x)$, we obtain a linear partial differential equation for the  continuous approximation of the average purity $P_c(t, x)$ (subscript $c$ refers to the continuous approximation),
\be\label{eq_s:pde_purity}
\partial_t P_c = \mu \partial_x^2P_c - \beta P_c, \qquad
P_c(t, x=0)= P_c(t, x=N) = 1.
\ee
For simplicity, we take a product initial state with $P_c(t=0, x)= 1$. 
Substituting $ P_c = 1+ \delta P_c$ we obtain
\be \label{eq:tilde_P}
\partial_t \delta P_c = \mu\partial_{xx} \delta P_c - \beta \delta P_c - \beta,
\ee
with zero initial and boundary conditions $\delta P_c(t, 0)= \delta P_c(t, N) = \delta P_c(0, x) = 0$.
Equation (\ref{eq:tilde_P}) can be solved by the Fourier method.
Expanding the solution in a Fourier series,
\be 
\delta P_c = \sum_{n=1}^{\infty} A_n(t) \sin{\left(k_n x \right)}, \qquad k_n = n\pi/N,
\ee
and expanding the constant term $\beta$ on the RHS of (\ref{eq:tilde_P}) in a Fourier series,
\be
\beta = \sum_{n=1}^{\infty} \beta_n \sin{k_n x}, \quad \text{where}\quad 
\beta_n=\frac{2\beta}{N}\int_0^{N}\sin{(k_n x)} dx=\begin{cases}
    0, & \text{if}\; n=2m\\
    4\beta/{\pi n }, & \text{if}\; n=2m+1
\end{cases},
\ee
we obtain a set of ordinary differential equations for coefficients $A_n(t)$,
\be \label{eq:dot_An}
\dot A_n(t) = -(\beta+\mu)A_n(t)  - \frac{4 \beta}{\pi n }\delta_{n, odd}.
\ee
Here $\delta_{n,m}$ is the Kronecker delta function.
The solution to (\ref{eq:dot_An}) reads
\be A_{n}(t)=
\begin{cases}
    A_{n}(0) e^{-(\beta+\mu k_{n}^2)t}, & \text{if} \quad n=\text{even}\\
    A_{n}(0) e^{-(\beta+\mu k_{n}^2)t}  +\frac{4 \beta}{ k_{n}(\beta+\mu k_{n}^2)}\left(e^{-(\beta+\mu k_{n}^2)t}-1\right), & \text{if} \quad n=\text{odd}
\end{cases}.
\ee
Finally, using the initial conditions  $A_{n}(0)=0$,
we obtain the solutions
\begin{align} \label{eq:solution_PDE}
P_c(t, x) &= 1 + \frac{4\beta}{N} \sum_{m=0}^{\infty}
\left(e^{-(\beta+\mu k_{2m+1}^2)t}-1\right)\frac{\sin{(k_{2m+1} x)}}{ k_{2m+1}(\beta+\mu k_{2m+1}^2)}, \\ S_2(t,x)&=-\log(P_c(t, x)).
\end{align}
As $t\to\infty$, the exponent in~\eqref{eq:solution_PDE} vanishes.
The corresponding stationary purity (i.e.\ at $t\to\infty$) obeys the following equation:
\be \label{eq:p_Gtationary}
\mu \partial_{xx}P-\beta P=0.
\ee
The solution of Eq.~(\ref{eq:p_Gtationary}) with the  boundary conditions in Eq. \eqref{eq_s:pde_purity} reads
\be \label{eq:pc3} 
P_c(t\to\infty, x) = \frac{\cosh{\left(\alpha\left(x-N/2\right)\right))}}{\cosh{\left(\alpha N/2\right)}},
\ee
where $\alpha=\sqrt{\beta/\mu}$. Equation (\ref{eq:pc3}) also follows directly from Eq.\ (\ref{eq:solution_PDE}). 
The asymptotic value of the entropy at $x=N/2$ read:
\be 
S_2(t\to\infty, x=N/2)=\log\left[\cosh{\left(\alpha N/2\right)}\right].
\ee
In the thermodynamic limit ($N\to\infty$),  
we obtain volume-law growth of the entropy $S_2(t\to\infty, x=N/2) = \alpha N/2 - \log{2}+O(\exp(-\alpha N)).$

We now derive parameters $\beta$ and $\mu$ in the continuous Equation~\eqref{eq:kpz_equation} from the underlying discrete evolution.
To find parameter $\beta$, we consider a homogeneous solution of the discrete model, $P_x(t)=P_{x+1}(t)$. In this case, after a single time step, one has $P_x(t+1) = 2c P_x(t)$. Similarly, for the entropy evolution, $S_2(t+1,x) = S_2(t,x)+\beta$. Taking $S_2(t,x) \approx -\log P_c(t,x)$ and comparing these two expressions, we find
\be
\beta = -\log(2c) = \log\Bigl(\frac{q^2+1}{2q}\Bigl).
\ee
To find the diffusion coefficient $\mu$, we consider a stationary solution $P_x(t) = P_x(t+1)$. Let us use the following ansatz for the solution to the discrete equation:
$
P_{x+1}(t) = \gamma P_x(t).
$
Then the condition of time-invariance leads us to
\be
P_x(t+1) = P_x(t)  = c\Bigl(\frac 1\gamma+\gamma\Bigl)P_x(t).
\ee
This condition is satisfied if
\be
\gamma = \frac{1}{2c}\Bigl(1\pm\sqrt{1-4c^2}\Bigl)\in\{q,q^{-1}\},
\ee
which corresponds to positive and negative slopes for the R\'{e}nyi entropy of the first and second halves of the chain, respectively. 

Assuming continuity of the entropy, we find that
\be
\frac{\partial S_2}{\partial x} \simeq S_2(t,x+1)-S_2(t,x) = \sqrt{\frac{\beta}{\mu}} = \log q.
\ee
This finally leads to the expression
\be
\mu = \frac{\log\left[(q^2+1)/2q\right]}{(\log q)^2}.
\ee
For $q=2$, this gives $\mu \approx 0.464$, close to the phenomenological value $\mu = 1/2$ considered in Ref.~\cite{EmergentStatNahum2019}. In the limit $q\to\infty,$ the diffusion coefficient vanishes as $\mu\sim 1/\log q$.

\textbf{Derivation of Eq.\ (\ref{eq:dt_G_0}):}
 Using the expression for the evolution of the Hartley entropy under unitary gates, $S_0(t+1, x)=\min{[S_0(t, x-1), S_0(t,x+1)]} + \log{q}$, and using identities $\min(a,b) = a + \min(b-a, 0)$ and $\min(a,0) = \frac{a-|a|}{2}$,  we find that
\be\label{eq:s0_sm}
\begin{split}
\partial_t S_0 \simeq  S_0(t+1, x)-S_0(t, x)\simeq 
\frac{[S_0(t, x+1) - S_0(t, x-1) - |S_0(t, x+1) - S_0(t, x-1)|]}{2} \\+ S_0(t, x-1) - S_0(t, x) + \log{q} \simeq -|\partial_x S_0| + \frac{1}{2}\partial_{x}^2S_0 + \log{q},
\end{split}
\ee
which results in Eq.~\eqref{eq:dt_G_0}. In deriving~\eqref{eq:s0_sm}, we used approximations $[S_0(t, x+1)-S_0(t, x-1)]/2\simeq \partial_x S_0$ and $S_0(t, x+1)-S_0(t, x) \simeq \partial_x S_0 + \frac{1}{2}\partial_{x}^2 S_0$.

\textbf{Solution of the hydrodynamic Equation~\eqref{eq:dt_G_0} for $S_0$ in 1D:}
In the case of a product initial state, the initial condition for the Hartley entropy in 1D is $S_0(t=0, x) = 0$. It is easy to see by direct substitution that the solution of the hydrodynamic equation for the Hartley entropy (see Eq.~\eqref{eq:dt_G_0} and Table~\ref{table:1}) reads:
\be \label{eq:SM_s0_solut}
    \frac{S_0(t, x)}{\log{q}} = 
    \begin{cases}
     x, \quad \textrm{if} \quad 0\leq x\leq t, \\
    t, \quad \textrm{if} \quad t\leq x\leq N -  t, \\
    t -  x, \quad \textrm{if} \quad N -  t \leq x\leq N. 
    \end{cases}
\ee
Moreover, within each
segment defined in Eq.\ (\ref{eq:SM_s0_solut}), the second derivative $\partial_{x}^2 S_0$ vanishes, except for special points where the second derivative is discontinuous. Therefore, in the special case when the initial state is a product state, the hydrodynamic equation for the Hartley entropy reduces to
\be \label{eq:SM_s0}
\partial_t \widetilde S_0  = 1 - |\partial_x \widetilde S_0|,
\ee
where $\widetilde S_0=S_0/\log{q}$.

On the other hand, we can write the hydrodynamic KPZ equation for the R\'{e}nyi entropy $\widetilde S_2=S_2/\log{q}$ in the scaling limit $q\to\infty$:
\be \label{eq:SM_s2}
\partial_t \widetilde S_2 \simeq - (\partial_x \widetilde S_2)^2 + 1,
\ee
where we used the asymptotic values for the coefficients $\beta = \log((q^2+1)/2q) \to \log{q}$ and $\mu = \beta/(\log{q})^2\to 1/\log{q}$ at $q\to \infty$.
Note that the piecewise solutions with slopes $0$, $\pm1$ shown in in Eq. (\ref{eq:SM_s0_solut}) satisfy both partial differential equations, Eq.~(\ref{eq:SM_s0}) and Eq.~(\ref{eq:SM_s2}). This analysis agrees with the statement made in the main text based on the Ising-model analogy: in the limit $q\to \infty$,
 R\'{e}nyi and Hartley entropies coincide, i.e.\ $S_0(t, x)\to S_2(t, x)$, provided that the initial state is a product state, i.e.\ $S_0(0, x)=S_2(0,x)=0$. 
 
\begin{figure}[t!]
    \includegraphics[width=1\textwidth]{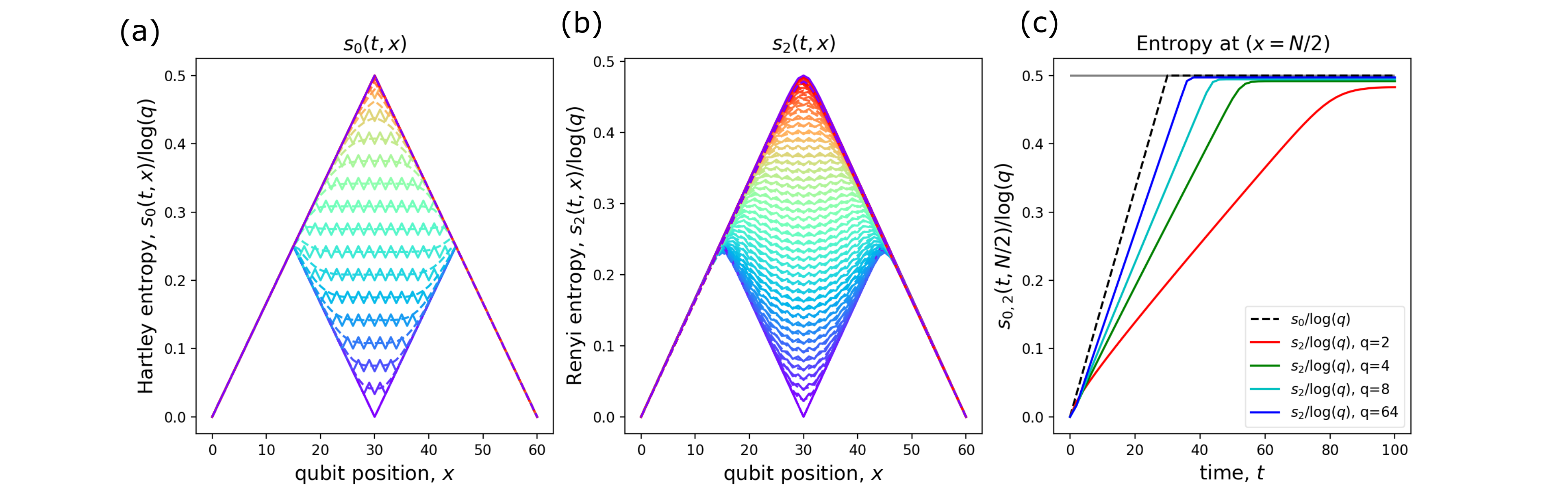}
    \caption{
    (a,b) Evolution of second R\'{e}nyi and Hartley entropies in 1D under a brickwork Haar-random unitary circuit: comparison of discrete-time evolution and the solutions of the corresponding continuous-time hydrodynamic equations (equations are shown in Table~\ref{table:1}). Here we plot the specific entropy (entropy per qudit) $s_{0,2}=S_{0,2}/N$ for $q=2$. The continuous-time equations were solved by a finite-difference scheme with a time-step $dt=10^{-2}$. 
    We chose an initial state where the two halves of the 1D system are disentangled from each other but each is in a maximally entangled state. 
    (c) Second R\'{e}nyi and Hartley entropies for a cut in the middle of the 1D qudit chain ($x=N/2$) as a function of time $t$ for $q = 2, 4, 8, 64$, calculated using discrete time-evolution equations. The Hartley entropy has a universal scaling law $s_0\sim \log{q}$ for arbitrary $q$. The $2$-R\'{e}nyi  entropy approaches the Hartley entropy in the limit of large qudit dimension $\log q \rightarrow \infty$: $s_2(t, x)\to s_0(t, x)$. At long times, both entanglement entropies saturate to a volume-law scaling in the thermodynamic limit: $S_0(t\to\infty, x=N/2) \to \frac{N}{2}\log{q}$ and $S_2(t\to\infty, x=N/2) \to \frac{N}{2}\log{q} - \log{2}$. }
    \label{figS3}
\end{figure}

 In Fig.\ \ref{figS3}, we compare the dynamics of second R\'{e}nyi and Hartley entropies in 1D by solving discrete (exact) and hydrodynamic (approximate) evolution equations listed in Table \ref{table:1}. In Fig.\ \ref{figS3}(a,b), one can see that the hydrodynamic evolution (dashed lines) agrees well with the exact solution (solid lines).
 At asymptotically large times, the stationary solutions for $s_0 = S_0/N$ and $s_2 = S_2/N$ have the same shape. In Fig.\ \ref{figS3}(c), we plot the values of the entropy in the middle of the 1D chain, $x=N/2$, for increasing values of the qudit dimension $q$. The plot shows that, in the asymptotic limit $q\to \infty$, the second R\'{e}nyi entropy approaches the Hartley entropy, $s_0(t,x)\to s_2(t, x)$.  
 
\begin{center}
\begin{table}[h!]
\caption{Discrete-time and continuous-time evolution equations for the second R\'{e}nyi entropy and for the Hartley entropy.}
\label{table:1}
\begin{tabular}{||c|c|c||}
 \hline
& R\'{e}nyi  entropy $S_2$  & Hartley entropy $S_0$ \\ [0.5ex] 
 \hline\hline
 Discrete-time evolution &
\parbox[t][1.6cm]{5cm}{ \(\displaystyle P^{t+1}_x=\frac{q}{q^2+1}(P_{x-1}^t+P_{x+1}^t),\)  \\ \vspace{.3cm}  \(\displaystyle S^{t+1}_{2,x} = -\log{P^{t+1}_x} \) }
 & \(\displaystyle S^{t+1}_{0,x}=\min{(S^t_{0, x-1}, S^t_{0, x+1})} + \log{q} \) \\ [0.5ex]
 \hline

  Hydrodynamic evolution & \(\displaystyle \partial_t S_2 = \mu \left(\partial_{x}^2 S_2 - (\partial_x S_2)^2 \right) + \beta \) & \parbox[t][.8cm]{6cm}{ \(\displaystyle \partial_t S_0 = \frac{1}{2} \partial_{x}^2 S_0 - |\partial_x S_0| + \log{q} \)} \\
 \hline
\end{tabular}
\end{table}
\end{center}

\section{Section IV: Mapping to a classical model in 1D}

In this section, we derive the classical Hamiltonian $H$ and the projector $\mathcal P$ in Eqs.~(\ref{eq:stat_model},\ref{eq:stat_model_ham}) in the main text. 

We consider combined circuit layers each consisting of a single unitary layer and a single measurement layer in reverse chronological order. This combined circuit can be mapped to a collection of trajectories $\mathcal G = \{G'_1\to G'_2\to \dots \to G'_n\}$ as described in the main text, where the Liouvillian $\mathcal L^{\rm u m}_k$ for a single combined layer connects set $G'_k$ to $G'_{k+1}$, and each trajectory contributes to the purity according to Eq.~\eqref{eq:Pdecomposition}. Our goal is now to map each trajectory to a classical state on a lattice with dynamical variables attached to the center of each lattice cell and to edges associated with measurements. At the end of this section, we cover open boundary conditions.

\begin{figure}[t!]
    \centering
    \includegraphics[width=1\textwidth]{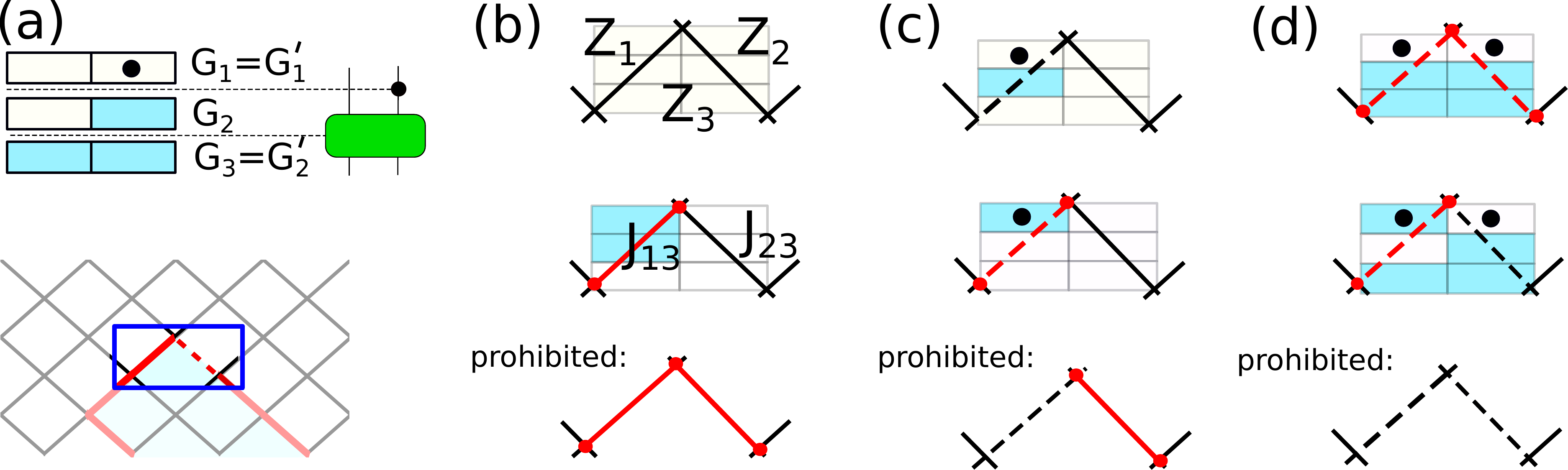}
    \caption{\textbf{Correspondence between trajectories $\mathcal G$ and configurations of a dynamical square lattice.} 
    (a) Top: two-qudit trajectory $\mathcal G = \{G_1 \to G_2\to G_3\} = \{\emptyset\to \Omega_2\to F\}$ illustrated by blue and white cells reflecting, respectively, whether the qudit belongs to $G_k$ or not. The monitored circuit generating such a trajectory is shown on the right (the black dot is a single-qudit measurement, while the green rectangle is a two-qudit gate). Bottom: the blue rectangle marks the portion of the square lattice corresponding to the $\{G_1 \to G_2\to G_3\}$ trajectory shown at the top. Here the colors of the top left and top right square cells reflect the colors of qudits in the initial configuration $G'_1 = G_1$, while the color of the bottom square reflects the color of the final configuration $G_2' = G_3$. The intermediate configuration $G_2$ is encoded in the edge below the measurement (solid edge if the cell keeps its color in $G_2$ and dashed edge if the color changes).
    (b-d) Illustration of the correspondence between trajectories (on the background) and lattice configurations (b) in the absence of measurements, (c) with a measurement of one qudit, and (d) with  measurements of both qudits. The examples of prohibited lattice configurations are shown at the bottom; these configurations are projected out using operator $\mathcal P$. }
    \label{figs:classical_model}
\end{figure}

As a toy illustrative example, let us focus on a primitive circuit containing only two qudits, labeled $\Omega_1$ and $\Omega_2$, and a single combined unitary-measurement layer applied to them [see top of Fig.~\ref{figs:classical_model}(a)]. This primitive system corresponds to a simple lattice with only three vertices and two edges shown inside a blue rectangle at the bottom of Fig.~\ref{figs:classical_model}(a). Therefore, our proposed classical model has three lattice cell variables ($Z_1$, $Z_2$, $Z_3$) and two edge variables ($J_{13}$ and $J_{23}$) [see labels in Fig.~\ref{figs:classical_model}(b)]. To study the mapping of trajectory $\{G_1'\to G_2'\}$ to a classical state on a lattice, we recall the trajectory defined in Fig.\ \ref{fig:1dcircuit}(b) of the main text, where each unitary and measurement layer is treated separately. Our single-step trajectory $\{G_1'\to G_2'\}$ thus corresponds to a two-step trajectory
$\{G_1\to G_2\to G_3\}$ in the language of Fig.\ \ref{fig:1dcircuit}(b). For both trajectories, the starting configuration is $G_1 = G'_1 = A$. Notably, as we will see below, the final state for any trajectory is either $G'_2 = G_3 = F$ (full system) or $\emptyset$, regardless of measurement locations. 

Let us start from the case of a unitary circuit without measurements. Such a circuit can be described by six possible trajectories including all possible initial states $A$, as shown below in the first column of Eq.\ (\ref{eqs:unitary_mapping}). These trajectories can be mapped to the classical spin configurations shown in the second column of Eq.\ (\ref{eqs:unitary_mapping}). The contribution $\mathcal P_{\mathcal G}$ of each trajectory ${\mathcal G}$ to the purity in Eq.~\eqref{eq:Pdecomposition} is given in the third column in the form of the configuration energy $E[\mathcal G]$:
\begin{align}\label{eqs:unitary_mapping}
&\text{Trajectory } \mathcal G  & \text{Classical configuration}\qquad\qquad\qquad\qquad & E[{\mathcal G}]=-\log(\mathcal P_{\mathcal G})/\beta \nonumber\\
&\{\emptyset\to\emptyset\to\emptyset\}, &\{Z_1=-1,Z_2=-1,Z_3=-1,J_{13}=1,J_{23}=1\}\qquad &\qquad0\nonumber\\
&\{\Omega_1\to\Omega_1\to\emptyset\},  &\{Z_1=+1,Z_2=-1,Z_3=-1,J_{13}=1,J_{23}=1\}\qquad &\qquad1\nonumber\\
&\{\Omega_1\to\Omega_1\to F\}, &\{Z_1=+1,Z_2=-1,Z_3=+1,J_{13}=1,J_{23}=1\}\qquad &\qquad1\nonumber\\
&\{\Omega_2\to\Omega_2\to \emptyset\}, &\{Z_1=-1,Z_2=+1,Z_3=-1,J_{13}=1,J_{23}=1\}\qquad &\qquad1\\
&\{\Omega_2\to\Omega_2\to F\}, &\{Z_1=-1,Z_2=+1,Z_3=+1,J_{13}=1,J_{23}=1\}\qquad &\qquad1\nonumber\\
&\{F\to F\to F\}, &\{Z_1=+1,Z_2=+1,Z_3=+1,J_{13}=1,J_{23}=1\}\qquad &\nonumber\qquad0
\end{align}

For illustration purposes, the first two trajectories in Eq.\ (\ref{eqs:unitary_mapping}) are shown as the top two diagrams in Fig.~\ref{figs:classical_model}(b). There, as in Fig.\ \ref{fig:1dcircuit}(b) in the main text, we mark qudits that belong to $G_k$ as blue and the rest of the qudits as white. For unitary circuits, the edges are not treated as dynamical variables; instead, the edges are all taken to be numbers equal to unity, i.e.\ $J_{ij}=+1$. Also, no trajectory is mapped to the configurations $\{Z_1=+1,Z_2=+1,Z_3=-1,J_{13}=1,J_{23}=1\}$ and $\{Z_1=-1,Z_2=-1,Z_3=+1,J_{13}=1,J_{23}=1\}$; therefore, these configurations must be projected out using the operator $\mathcal P$. We show the first of these prohibited configurations at the bottom of Fig.~\ref{figs:classical_model}(b). 
Each allowed configuration contributes with energy $E[\mathcal G]=1$ equal to the total length of domain walls (i.e.\ the number of red edges).

Now let us assume that a measurement is applied to one of the qudits $\Omega_k$, where $k=1$ or $k = 2$, while the other qudit $\Omega_{\overline{k}}$ remains unmeasured (here we define $\overline{1}=2$ and $\overline{2} = 1$). The measurement adds more accessible trajectories to the ones already listed in  Eq.~\eqref{eqs:unitary_mapping}. Here are these additional trajectories, the corresponding classical configurations, and the corresponding energies:
\begin{align}\label{eqs:single_meas_mapping}
&\text{Trajectory } \mathcal G  & \text{Classical configuration}\qquad\qquad\qquad\qquad & E[{\mathcal G}]=-\log(\mathcal P_{\mathcal G})/\beta \nonumber\\
&\{\emptyset\to\Omega_k\to\emptyset\}, &\{Z_k=-1,Z_{\overline{k}}=-1,Z_3=-1,J_{k3}=-1,J_{\overline{k}3}=1\}\qquad &\qquad1\nonumber\\
&\{\emptyset\to\Omega_k\to F\},  &\{Z_k=-1,Z_{\overline{k}}=-1,Z_3=+1,J_{k3}=-1,J_{\overline{k}3}=1\}\qquad &\qquad1\nonumber\\
&\{\Omega_k\to\emptyset\to \emptyset\}, &\{Z_k=+1,Z_{\overline{k}}=-1,Z_3=-1,J_{k3}=-1,J_{\overline{k}3}=1\}\qquad &\qquad0\nonumber\\
&\{\Omega_{\overline{k}}\to F\to F\}, &\{Z_k=-1,Z_{\overline{k}}=+1,Z_3=+1,J_{k3}=-1,J_{\overline{k}3}=1\}\qquad &\qquad0\\
&\{F\to\Omega_{\overline{k}}\to F\}, &\{Z_k=+1,Z_{\overline{k}}=+1,Z_3=+1,J_{k3}=-1,J_{\overline{k}3}=1\}\qquad &\qquad1\nonumber\\
&\{F\to\Omega_{\overline{k}}\to \emptyset\}, &\{Z_k=+1,Z_{\overline{k}}=+1,Z_3=-1,J_{k3}=-1,J_{\overline{k}3}=1\}\qquad &\nonumber\qquad1
\end{align}
Trajectories 2 and 3 are shown as the top two diagrams in Fig.~\ref{figs:classical_model}(c) for $k=1$ (and $\overline{k}=2$). As for the unitary case, $J_{\overline k,3} = 1$; however, since we measured $\Omega_k$, $J_{k,3}$ is now a dynamical variable. Since no trajectories correspond to  configurations $\{Z_1=-1,Z_2=+1,Z_3=-1,J_{k3}=-1,J_{\overline{k}3}=1\}$ and $\{Z_1=+1,Z_2=-1,Z_3=+1,J_{k3}=-1,J_{\overline{k}3}=1\}$, these configurations must be projected out by $\mathcal P$. The first of these two prohibited configurations is shown at the bottom of Fig.~\ref{figs:classical_model}(c).

Finally, if both qudits are measured, the following trajectories become accessible, in addition to those already listed in Eq.~\eqref{eqs:unitary_mapping} and Eq.~\eqref{eqs:single_meas_mapping}:
\begin{align}\label{eqs:two_meas_mapping}
&\text{Trajectory } \mathcal G  & \text{Classical configurations}\qquad\qquad\qquad\qquad & E[{\mathcal G}]=-\log(\mathcal P_{\mathcal G})/\beta \nonumber\\
&\{\emptyset\to F \to F\}, &\{Z_1=-1,Z_2=-1,Z_3=+1,J_{13}=-1,J_{23}=-1\}\qquad &\qquad0\nonumber\\
&\{\Omega_1\to\Omega_2\to\emptyset\},  &\{Z_1=+1,Z_2=-1,Z_3=-1,J_{13}=-1,J_{23}=-1\}\qquad &\qquad1\nonumber\\
&\{\Omega_1\to\Omega_2\to F\}, &\{Z_1=+1,Z_2=-1,Z_3=+1,J_{13}=-1,J_{23}=-1\}\qquad &\qquad1\nonumber\\
&\{\Omega_2\to\Omega_1\to\emptyset\}, &\{Z_1=-1,Z_2=+1,Z_3=-1,J_{13}=-1,J_{23}=-1\}\qquad &\qquad1\\
&\{\Omega_2\to\Omega_1\to F\}, &\{Z_1=-1,Z_2=+1,Z_3=+1,J_{13}=-1,J_{23}=-1\}\qquad &\qquad1\nonumber\\
&\{F\to \emptyset\to \emptyset\}, &\{Z_1=+1,Z_2=+1,Z_3=-1,J_{13}=-1,J_{23}=-1\}\qquad &\nonumber\qquad0
\end{align}
Trajectories 1 and 3 are shown as the top two diagrams in Fig.~\ref{figs:classical_model}(d). Since no trajectories map to configurations $\{Z_1=+1,Z_2=+1,Z_3=+1,J_{k3}=-1,J_{\overline{k}3}=-1\}$ and $\{Z_1=-1,Z_2=-1,Z_3=-1,J_{k3}=-1,J_{\overline{k}3}=-1\}$, these configurations must be projected out by $\mathcal P$. Both prohibited configurations correspond to the ``lambda''-configuration of dashed lines shown at the bottom of Fig.~\ref{figs:classical_model}(d).

Using Eqs.~\eqref{eqs:unitary_mapping}, \eqref{eqs:single_meas_mapping}, and \eqref{eqs:two_meas_mapping}, it is straightforward to verify that the energy of each allowed configuration is described by the Ising-type interaction,
\be
E[\mathcal G] = 1-\frac 12\Bigl(J_{13}Z_1Z_3+J_{23}Z_2Z_3\Bigl),
\ee
while all prohibited configurations satisfy
\be
\frac{1}{4}(Z_3-J_{13}Z_1)(Z_3-J_{23}Z_2)=1.
\ee
Therefore, the purity of set $A$ is given by
\be
\tilde P_A = \Tr \Bigl(\mathcal P\exp(-\beta H)\Bigl), \qquad H = 1-\frac 14\sum_{\<i,j\>}J_{ij}Z_iZ_j,
\ee
 where the projector $\mathcal P$ excludes prohibited configurations as well as sets the initial conditions for trajectory $\mathcal G$:
\be
\mathcal P = \mathcal P_A\Bigl(1-\frac 14(Z_3-J_{13}Z_1)(Z_3-J_{23}Z_2)\Bigl),
\ee
where $\mathcal P_A = \frac 14 \prod_{i\in A}(1-Z_i)\prod_{j\in F\setminus A}(1+Z_j)$, where products are taken over the $Z$-variables at the top of the lattice (i.e.\ $Z_1$ and $Z_2$).

Having considered the case of two qudits and one layer, we can generalize the result to an arbitrary number of qudits and layers. This generalization is possible because: (a) the process is Markovian, so contributions from different layers do not interfere; (b) gates in the same layer do not overlap, so their contributions are also independent. Therefore, the full multi-qudit multi-layer configuration on a square lattice is obtained by glueing together 2-qudit 1-layer elements described above, while the effective energy of this full configuration is just the sum of the energies of these elements:
\be
H = \frac{N_0}2-\frac 14\sum_{\<i,j\>}J_{ij}Z_iZ_j, \qquad \mathcal P = \mathcal P_A \prod_{(i,j,k)\in\Sigma_3}\Bigl(1-\frac 14 (Z_k-J_{ik}Z_i)(Z_k-J_{jk}Z_j)\Bigl),
\ee
which is Eq.~\eqref{eq:stat_model} in the main text.

Another approach to dealing with prohibited configurations (without projecting them out) is to consider a modified Ising Hamiltonian with a penalty term:
\be
\tilde P_A = \lim_{\Gamma\to\infty} \Tr \mathcal P_A\exp(-\beta H_\Gamma),
\ee
where
\be
H_\Gamma = \frac 12 N_0-\frac 14\sum_{\<i>j\>}J_{ij}Z_iZ_j+ \Gamma\sum_{(i,j,k)\in\Sigma_3}(Z_k-J_{ik}Z_i)(Z_k-J_{jk}Z_j).
\ee
In the limit of infinite penalty coefficient $\Gamma$,  the two approaches are equivalent.

One can compute the energies of domain-wall configurations without explicitly using the Hamiltonian. For unitary evolution, all Ising couplings are $J_{ij} = 1$;  therefore, the energy of any allowed Ising configuration $E[\mathcal G]$ is equal to the total length of domain walls. 
In the presence of measurements,
$E[\mathcal G]$ is equal to the total length of solid segments of domain walls (i.e.\ solid red edges) plus the total number of dashed edges outside of the domain walls (i.e.\ dashed black edges). The projector $\mathcal P$ prohibits  configurations shown in Fig.~\ref{fig:1dcircuit}(c). Because dashed edges resemble cuts on a lattice, we can draw an analogy of the model to trajectories on a \textit{dynamical} percolated lattice.

\begin{figure}[t!]
    \centering
    \includegraphics[width=0.4\textwidth]{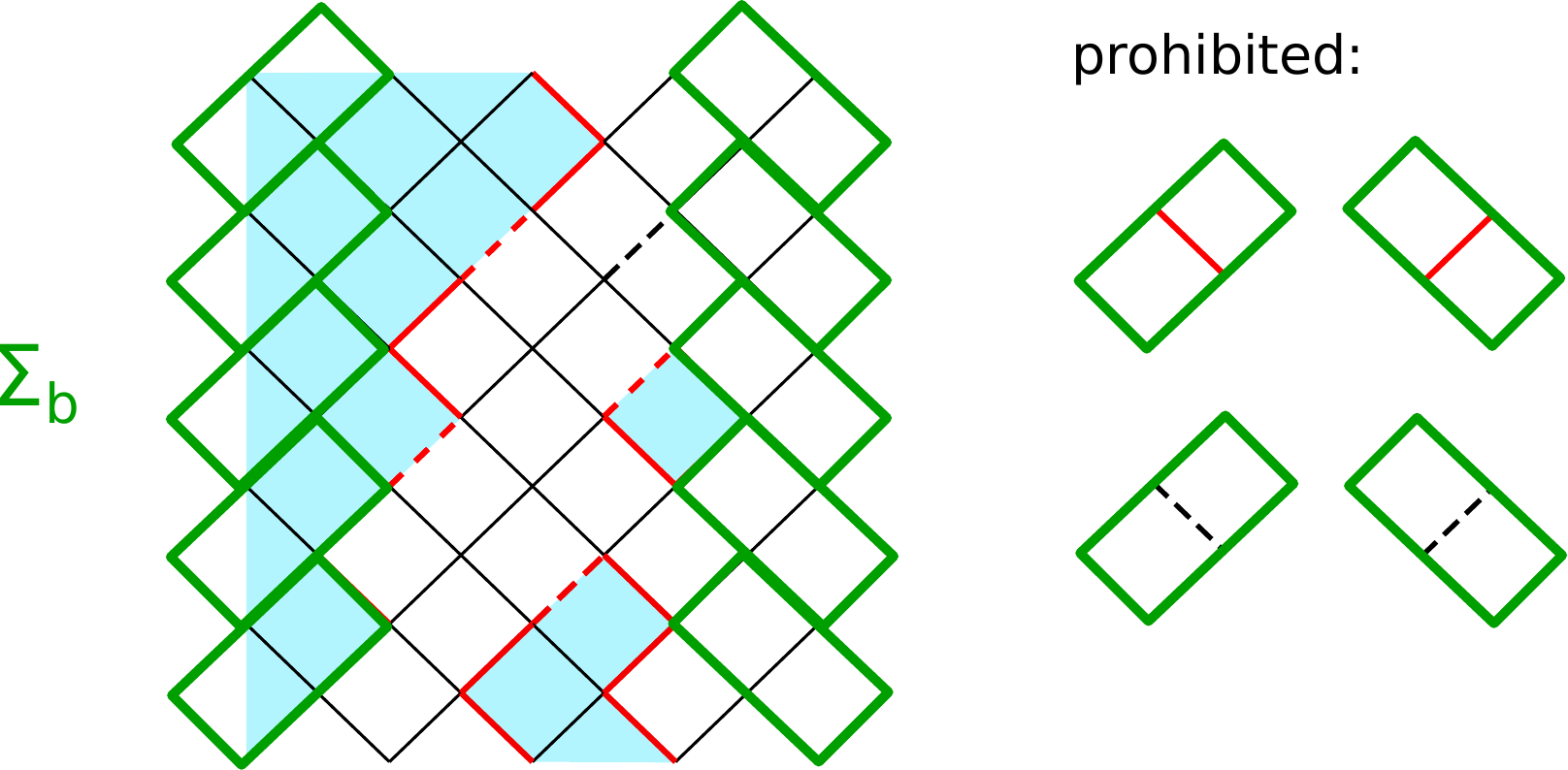}
    \caption{\textbf{Boundary conditions}. Open boundaries put additional constraints on the allowed spin-model configurations. Edge pairs of squares (green boxes) cannot exhibit configurations shown on the right, i.e.\ solid red and dashed black edges are prohibited in the indicated locations independently of square colors.}
    \label{figs:bc_spin_model}
\end{figure}

\textbf{Boundary conditions}. Boundaries result in the presence of unpaired qudits/cells at the edges that are not affected by the unitaries at even (or odd, depending on where exactly the boundary is) layers, as can be seen from Fig.~\ref{fig:1dcircuit}(a). These qudits exhibit additional prohibited trajectories at even (or odd) times. Specifically, black dashed and solid red lines are prohibited in the locations shown in Fig.~\ref{figs:bc_spin_model}. 
Such trajectories can be projected out by taking $\mathcal P \to  \mathcal P\prod_{ij\in \Sigma_b}\frac 12(1+J_{ij}Z_iZ_j)$ for all boundary square pairs $\Sigma_b$ highlighted as green boxes in Fig.~\ref{figs:bc_spin_model}. 

\section{SECTION V: Continuous evolution}
\label{app:contmeas}

In this section, we show how continuous evolution of the system under a stochastic Hamiltonian and local continuous measurements gives rise to Eqs.~\eqref{eq:cont_unit_evo} and \eqref{eq:cont_meas_main} in the main text. We also analyze the accuracy of the mean-field approximation used to simplify Eq.\ (\ref{eq:cont_meas_main}). Finally, we show that the continuous evolution in  Eqs.~(\ref{eq:cont_unit_evo},\ref{eq:cont_meas_main}) can be used to derive the discrete evolution in  Eqs.~(\ref{eq:unitary_purity_evo},\ref{eq:measurement_evo})

First, we prove that the eigenbasis of the monitored operators can be randomized while preserving purities. This randomization will allow us to derive a differential equation for the purities.
 More specifically, we consider random time-dependent unitaries $U^\dag_{\tau,i}(t)$ supported on a single qudit $i$ and drawn from the circular unitary ensemble. 
 In order to have a time-continuous description, we assume that the matrices are weakly correlated in time, $\<[U_{\tau,i}(t)]_{\mu\nu}[U^*_{\tau,j}(t')]_{\mu'\nu'}\> = q^{-1}\delta_{\mu\mu'}\delta_{\nu\nu'}\delta_{ij}f[(t-t')/\tau]$, where $f[x]$ is a smooth correlation function satisfying $f[0]=1$ and $\lim_{|x|\to\infty}f[x]=0$. The correlation $f(x)$ will be later removed by taking the limit $\tau\to0$. 
 In this limit, the unitaries represent a stochastic process where, at each time, the matrix $U_{\tau=0,i}(t)$ is drawn independently from the circular unitary ensemble.

Consider the unitary $U_{\tau}(t) = \prod_i U_{\tau,i}(t)$ and the corresponding rotating-frame wavefunction
\be\label{eqs:randomize}
\begin{split}
|\psi'(t)\> = U^\dag_{\tau}(t)|\psi(t)\>.
\end{split}
\ee
Since $U_{\tau,i}(t)$ are defined as single-qudit unitary operators, the purity $P_G$ for any $G$ remains invariant under this transformation:
\be\label{eqs:randomizePG}
\begin{split}
P_G(|\psi'\>\<\psi'|) = P_G(|\psi\>\<\psi|).
\end{split}
\ee
The new wavefunction $|\psi'(t)\>$ satisfies 
\begin{align}
\label{eq-app:rho_monit_master}
\frac{d}{dt} \ket{\psi'} = -i\Bigl(H'(t)+H_U(t)\Bigl)|\psi'\>+\left[ -\sum_j \kappa_j \Big( O_j(t) - \bra{\psi'} O_j(t) \ket{\psi'} \Big)^2 + \sum_j  \xi_j(t)\sqrt{2 \kappa_j} \Big( O_j(t) - \bra{\psi'} O_j(t) \ket{\psi'} \Big) \right] \ket{\psi'},
\end{align}
where
\begin{eqnarray}
H_U(t) &\equiv& -i\sum_i U^\dag_{\tau,i}(t)\partial_tU_{\tau,i}(t),\\
H'(t) &\equiv& U^\dag_{\tau}(t)H(t)U_{\tau}(t) = \sum_i I_{F\setminus\omega_i}\otimes h'_i(t),\\
O_j(t) &\equiv& U^\dag_{\tau}(t)O_jU_{\tau}(t) = U^\dag_{\tau,j}(t)O_jU_{\tau,j}(t).
\end{eqnarray}
The modified stochastic Hamiltonians $h_i'(t) \equiv U^\dag_{\tau,i}(t)h_i(t)U_{\tau,i}(t)$ have the same distribution as $h_i(t)$, due to the invariance of the Gaussian unitary ensemble 
under unitary rotations. The new Hamiltonian term $H_U(t)$ is a sum of single-qudit operators.  Therefore, the rotation in Eq.~\eqref{eqs:randomize} effectively randomizes the measurement bases: $O_j \to U^\dag_{\tau,j}(t) O_j U_{\tau,j}(t)$. The penalty for this is the addition of single-qudit terms to the Hamiltonian, but with no effect on the distribution of the original stochastic Hamiltonian terms.  Importantly, we show below that the additional single-qudit terms do not affect (averaged)  purities $P_G$.

\textbf{Unitary dynamics}. Let us first focus on the effect of the combined Hamiltonian
\be
H_c(t) = H_U(t)+\sum_iH'_i(t),
\ee
where each local stochastic term can be represented as $H'_i(t) = \mathbb I_{F\setminus\omega_i}\otimes h'_i(t)$, where $h'_i(t)$ 
are independent random matrices supported on sets $\omega_i$ as introduced in the main text.
The unitary evolution operator for infinitesimal time $dt$ is
\be
U_{t,dt} = \mathcal T\exp\Bigl(-i\int_t^{t+dt} dt'H_c(t')\Bigl).
\ee
Therefore, the evolution of the density matrix satisfies
\be
\begin{split}
\rho(t+dt) = U_{t,dt}\rho U^\dag_{t,dt} =& \rho(t)-i\int_{t}^{t+dt}[H_U(t'),\rho]dt' -i\sum_i\int_{t}^{t+dt}[H'_i(t'),\rho]dt'\\
&-\frac 12\sum_{ij}\int_t^{t+dt}dt'\int_{t}^{t+dt}dt''[H'_i(t'),[H'_j(t''),\rho]]+O(dt^2).
\end{split}
\ee
The evolution of purity is connected to the density matrix $\rho = |\psi'\>\<\psi'|$ as follows:
\be \label{eq:SM_PG_tr}
\frac{d}{dt} P_G = \frac{d}{dt}\<\Tr_G\rho_G^2\> = \lim_{dt \rightarrow 0} \Bigl\<\Tr_G\frac{\rho_G^2(t+dt)-\rho_G^2(t)}{dt}\Bigl\>_{{\rm circ}, H'_i(t)}.
\ee
The averaging in 
Eq.\ (\ref{eq:SM_PG_tr}) includes both the circuit average $\langle \ldots \rangle_{\rm circ}$ over
different evolution histories realized by the stochastic Hamiltonian $H'_i$ prior to time $t$ and the average $\langle \ldots \rangle_{H_i'(t)}$ over the stochastic Hamiltonian $H'_i$ at the current time $t$. At this point, we are not yet averaging over $U_{\tau,i}$.

To calculate the quantity on the RHS of Eq.\ (\ref{eq:SM_PG_tr}), we use the reduction
\be
\<H'_i(t) O H'_j(t')\>_{H_i} = \frac 12\alpha_i \delta_{ij}\delta(t-t')\<W_i O W_i\>_W,
\ee
where $O$ is any operator and $W_i$ is a Wigner random matrix from a Gaussian unitary ensemble 
supported on $\omega_i$ and satisfying
\be
\<W_{ai,bj}W^*_{a'i',b'j'}\>_W = \frac 1{d_\omega}\delta_{aa'}\delta_{bb'}\delta_{ii'}\delta_{jj'}.
\ee
Since $\langle H_i'(t) \rangle_{H_i} =0$,
Eq.\ (\ref{eq:SM_PG_tr}) 
takes the form
\begin{align}
\label{eq:SM_dt_PG}
\frac{d}{dt} P_G &= -i\Tr_G\Bigl(\{\Tr_{F\setminus G}[H_U,\rho],\rho_G\}\Bigl)-\frac{\alpha_i}2\left\<\Bigl(\Tr_G\Bigl[ (\Tr_{F\setminus G}[W_i,\rho])^2\Bigl]+\Tr_G\Bigl( \Tr_{F\setminus G}[W_i,[W_i,\rho]\rho_G\Bigl)\Bigl)\right\>_{{\rm circ}, W}\nonumber\\
&= -\frac{\alpha_i}2 \Bigl\<\Bigl(\Tr_G \Bigl(\Tr_{F\setminus G}(W_i\rho)\Tr_{F\setminus G}(W_i\rho)\Bigl)+\Tr_G \Bigl(\Tr_{F\setminus G}(\rho W_i)\Tr_{F\setminus G}(\rho W_i)\Bigl) -2\Tr_G\Bigl(\Tr_{F\setminus G}(\rho W_i)\Tr_{F\setminus G}(W_i\rho )\Bigl)\nonumber\\
&+ \Tr_G \Bigl( \Tr_{F\setminus G}(W^2_i\rho) \rho_G\Bigl)+\Tr_G \Bigl( \Tr_{F\setminus G} (\rho W^2_i) \rho_G\Bigl) - 2\Tr_G\Bigl(\Tr_{F\setminus G} (W_i\rho W_i) \rho_G\Bigl)\Bigl)\Bigl\>_{{\rm circ}, W}.
\end{align}
The first term in \eqref{eq:SM_dt_PG} 
that contains $H_U$ vanishes since it acts on single qudits.

Averages in Eq. (\ref{eq:SM_dt_PG}) can be calculated using the following identities:
\be\label{eq:wigner_ham_details}
\begin{split}
&\Bigl\<\Tr_G \bigl(\Tr_{F\setminus G}(W\rho)\bigl)^2\bigl\>_W  = \Bigl\<\Tr_G \bigl(\Tr_{F\setminus G}(\rho W)\bigl)^2\bigl\>_W = \frac 1{d_\omega}\Tr_{G\Delta\omega}\Bigl(\rho_{G\Delta\omega}^2\Bigl),\\
&\Bigl\<\Tr_G
\Bigl(\Tr_{F\setminus G}(\rho W)\Tr_{F\setminus G}(W\rho)\Bigl)\Bigl\>_W = \frac 1{d_{\omega\setminus G}} \Tr_{G\cup\omega}(\rho^2_{G\cup\omega}),\\
&\Bigl\<\Tr_G \Bigl( \Tr_{F\setminus G}(W^2\rho) \rho_G\Bigl)\Bigl\>_W = \Bigl\<\Tr_G \Bigl( \Tr_{F\setminus G}( \rho W^2) \rho_G\Bigl)\Bigl\>_W = \Tr_{G}(\rho_{G}^2),\\
&\Bigl\<\Tr_G\Bigl(\Tr_{F\setminus G} (W\rho W) \rho_G\Bigl)\Bigl\>_W = \frac1{ d_{G\cap\omega}} \Tr_{G\setminus\omega}(\rho^2_{G\setminus\omega}).
\end{split}
\ee
Combining equations (\ref{eq:SM_dt_PG}) and (\ref{eq:wigner_ham_details}), we obtain
\be\label{eqs:unitary_part}
 \frac{d}{dt} P_G = -\sum_i \alpha_i\Bigl(P_G+\frac {P_{G\Delta \omega_i}}{d_{\omega_i}}-\frac {P_{G\setminus\omega_i}}{d_{\omega_i\cap G}} -\frac {P_{G\cup\omega_i}}{d_{\omega_i\setminus G}} \Bigl), 
\ee
which is Eq.~\eqref{eq:cont_unit_evo} in the main text.

\textbf{Continuous monitoring}. Let us consider now the dynamics of the system under continuous measurements of observables $O_j(t)$ applied to  qudits $\Omega_j$ in the randomized basis,  $O_j(t) = U_{\tau,j}(t)O_jU^\dag_{\tau,j}(t)$: 
\begin{align}
\label{eq-app:rho_monit_master}
\frac{d}{dt} \ket{\psi'} = \left[ -\sum_j \kappa_j \Big( O_j(t) - \bra{\psi'} O_j(t) \ket{\psi'} \Big)^2 + \sum_j  \xi_j(t)\sqrt{2 \kappa_j} \Big( O_j(t) - \bra{\psi'} O_j(t) \ket{\psi'} \Big) \right] \ket{\psi'}.
\end{align}
Here coefficients $\kappa_j$ characterize the strength of the coupling to the measurement apparatus, and $\xi_j(t)$ are independent unbiased Gaussian random variables satisfying $\<\xi_i(t)\xi_j(t')\>_\xi = \delta_{ij}\delta(t-t')$. The average $\langle \dots \rangle_\xi$ over the random variables $\xi$ amounts to averaging over the measurement back-action in the realization of the monitoring. 
 We find that, upon averaging over $\xi_j$,  the purities evolve as
\begin{align}\label{eq:SM_P_G}
\frac{d}{dt} P_G = 2 \sum_{j} \kappa_j \Bigl\langle \Tr_G [\rho_G ,O_j(t)]^2 \Bigl\rangle_{{\rm circ}} + 2 \sum_j \kappa_j  \left\langle \Tr_{G}\Bigl[ \Tr_{F\setminus G}\bigl( \left\{O_j(t),\rho  \right\} - 2\tr{\rho  O_j(t)} \rho\bigl)   \Bigl]^2 \right\rangle_{{\rm circ}},
\end{align}
where, as defined above, the circuit avareage $\langle \ldots \rangle_{\rm circ}$ is over
the stochastic Hamiltonian $H'_i$ prior to time $t$. 

Equation (\ref{eq:SM_P_G}) can now be rewritten using the following identities (to ease notation, we suppress time dependence of $O_j$): 
\be
\begin{split}
&\Tr_G ([\rho_G,O_j]^2)  = - 2\Bigl( \Tr_G (\rho^2_G O^2_j)-\Tr_G (\rho_G O_j)^2\Bigl) \delta_{\Omega_j\in G},\\
& \trs{G}{ \Tr_{F\setminus G} \left\{\rho,O_j  \right\}}^2 = 2\Bigl( \Tr_G (\rho^2_G O^2_j)+\Tr_G (\rho_G O_j)^2\Bigl)\delta_{\Omega_j\in G}+4\trs{G}{ \Tr_{F\setminus G} (\rho O_j) }^2\delta_{\Omega_j\in {F\setminus G}},\\
& \trs{G} { \trs{{F\setminus G}}{ \left\{\rho,O_j  \right\} } \rho_G } =  2\trs{G} {\rho^2_G O_j }\delta_{\Omega_j\in G}+2\trs{L} { \trs{{F\setminus G}}{ \rho O_j  } \rho_G }\delta_{\Omega_j\in {F\setminus G}}.
\end{split}
\ee
where $\delta_{\Omega\in G} = 1$ if $\Omega\in G$ and $\delta_{\Omega\in G} = 0$ otherwise.

Since the choice of single-qudit rotations $U_\tau$ does not affect the purities [see Eq.\ (\ref{eqs:randomizePG})], we can additionally average the purities over $U_\tau$ without loss of generality.
It is also convenient to split the Liouvillian operator into 
two parts as
\begin{align}\label{eqs:monitoring_part_divided}
\frac{d}{dt} P_G = \sum_{j:\Omega_j\in G} \kappa_jM_1(\rho, O_j,G)+\sum_{j:\Omega_j\in F\setminus G} \kappa_j M_2(\rho,O_j,G),
 \end{align}
 where $M_1$ and $M_2$ correspond, respectively, to measurements inside and outside the set $G$:
 \begin{align}
 &M_1(\rho, O_j,G)= 8 \left\langle \Tr_G (\rho_G O_j)^2-2\Tr \rho O_j\trs{G} {\rho^2_G O_j }+(\Tr \rho O_j)^2\Tr_G\rho_G^2 \right\rangle_{{\rm circ},U_\tau}, \\
 &M_2(\rho,O_j,G) = 8  \left\langle \trs{G}{ \Tr_{F\setminus G} (\rho O_j) }^2-2\Tr \rho O_j\trs{G} { \trs{{F\setminus G}}{ \rho O_j  } \rho_G }+(\Tr \rho O_j)^2\Tr_G\rho_G^2 \right\rangle_{{\rm circ},U_\tau}  . 
\end{align}

In the limit of vanishing correlation time, $\kappa_j\tau\to0$, the unitaries $U_{\tau,i}(t)$ at each time $t$ can be considered as independent Haar-random matrices. Therefore, the average 
of the density operator $\rho$ over $U_{\tau,i}$ and the average of observable $O_j(t)$ at time $t$ over $U_{\tau,i}(t)$ can be performed independently of each other. 

Then, after averaging $O_j(t)$ over $U_{\tau,i}(t)$, we obtain
\begin{align}\label{eq:M1}
  M_1(\rho, O_j,G) &= \frac{8\Tr_{\Omega_j} O_j^2}{q^2-1}\left\langle \Tr_{G\setminus\Omega_j} \rho_{G\setminus\Omega_j}^2+\Tr_G\rho_G^2\Tr_{\Omega_j}\rho_{\Omega_j}^2 - 2\Tr_G(\rho_{\Omega_j}\rho^2_G) \right\rangle_{{\rm circ},U_\tau} , \\
M_2(\rho,O_j,G) &= \frac{8\Tr_{\Omega_j} O_j^2}{q^2-1} \left\langle \Tr_{G\cup\Omega_j} \rho_{G\cup\Omega_j}^2+\Tr_G\rho_G^2\Tr_{\Omega_j}\rho_{\Omega_j}^2 - 2\Tr_{G\cup\Omega_j}(\rho_{\Omega_j}\rho_G\rho_{G\cup\Omega_j}) \right\rangle_{{\rm circ},U_\tau}  \label{eq:M2}
.
\end{align}

We notice that $M_2$ can be written in a compact way,
\be\label{eqs:M2_simple}
 M_2(\rho,O_j,G) = \frac{8\Tr_{\Omega_j} O_j^2}{q^2-1}\left\langle  \big\| \rho_{G\cup\Omega_j} - \rho_{G} \otimes \rho_{\Omega_j} \big\|_2^2 \right\rangle_{{\rm circ},U_\tau} .
\ee
where $\|\cdot\|_2$ is a 2-norm.

Given that the overall state of the system is pure, we can also simplify $M_1$ in Eq.~\eqref{eq:M1} using the property
\be \label{eq:gfg}
\frac{d}{dt} P_G =\frac{d}{dt} P_{F\setminus G}.
\ee
Since variables $\kappa_j$ are independent variables that can take arbitrary values, Eq.~\eqref{eqs:monitoring_part_divided} will satisfy Eq.\ (\ref{eq:gfg}) only if
\be \label{eq:m1m2}
M_1(\rho,O_j,G) = M_2(\rho,O_j,F\setminus G).
\ee
Combining Eqs.\ (\ref{eqs:M2_simple}) and (\ref{eq:m1m2}), we obtain
 \be
  M_1 (\rho, O_j,G)=  \frac{8\Tr_{\Omega_j} O_j^2}{q^2-1}\left\langle \big\| \rho_{{(F\setminus G)}\cup\Omega_j} - \rho_{{F\setminus G}} \otimes \rho_{\Omega_j} \big\|_2^2 \right\rangle_{{\rm circ},U_\tau} .
 \ee

Therefore, the following 
expression holds for the combined evolution under Hamiltonian and measurement dynamics:
\begin{align}
 \frac {d}{d t}   P_G   &= -\sum_i \alpha_i\Bigl(P_G+\frac {P_{G\Delta \omega_i}}{d_\omega}-\frac {P_{G\setminus\omega_i}}{d_{\omega_i\cap G}} -\frac {P_{G\cup\omega_i}}{d_{\omega_i\setminus G}} \Bigl)  \\
 &+\sum_{j:\Omega_j \in  G}\zeta_j  \left\langle \big\| \rho_{\Omega_j \cup ({F\setminus G}) } -  \rho_{\Omega_j} \otimes \rho_{{F\setminus G}}  \big\|_2^2 \right\rangle_{\rm circ} \nonumber + \sum_{j: \Omega_j\in  {F\setminus G}} \zeta_j \left\langle  \big\| \rho_{G\cup\Omega_j} - \rho_{G} \otimes \rho_{\Omega_j} \big\|_2^2 \right\rangle_{\rm circ}, 
 \end{align}
where $\zeta_j =  8\Tr_{\Omega_j} O_j^2\kappa_j/(q^2-1)$, and we removed the average over $U_\tau$ in the expressions in the second line since these
expressions, like purity, are invariant under the $U_\tau$ rotation 
defined in Eq.~\eqref{eqs:randomize}. The second line of this equation 
yields Eq.~\eqref{eq:cont_meas_main} in the main text.

 In the mean-field approximation, the correlation distance can be approximated as
 \be\label{eqs:mf_corr_dis}
 \left\langle  \big\| \rho_{G\cup\Omega} - \rho_{G} \otimes \rho_{\Omega} \big\|_2^2 \right\rangle_{\rm circ} \simeq P_{G\cup\Omega}-P_GP_\Omega.
 \ee
The accuracy of this approximation is illustrated in Fig.~\ref{figS2}(c) for a 9-qubit system in a random matrix-product state, similar to Eq.~\eqref{eqs:annealed_approx}. As in the case of the annealed approximation, the error of the mean-field approximation strongly depends on the bond dimension $N_b$ and rapidly decreases from $\epsilon \sim 1$ for $\log_2 N_b =1$ down to $\epsilon \sim 10^{-2}$ for $\log_2 N_b =5$. 

\textbf{Relation to the discrete case}. Let us first derive Eq.~\eqref{eq:unitary_purity_evo} in the main text from Eq.~\eqref{eq:cont_unit_evo}.  Consider a situation where a single stochastic Hamiltonian term $h$ with strength $\alpha$ and support $\omega$ is applied to the system.
The purities $P_{G\setminus\omega}$ and $P_{G\cup\omega}$ remain unchanged throughout the evolution. Therefore, the evolution of purity $P_G$ can be evaluated by solving the following closed set of equations: 
\begin{eqnarray}
\frac{d}{dt}P_{G} &=& -\alpha\Bigl(P_{G}+\frac{P_{G\Delta\omega}}{d_{\omega}}-\frac{P_{G\setminus\omega}}{d_{G\cap\omega}}-\frac{P_{G\cup\omega}}{d_{\omega\setminus G}}\Bigl),\\
\frac{d}{dt} P_{G\Delta\omega}&=& -\alpha\Bigl(P_{G\Delta\omega}+\frac{P_{G}}{d_{\omega}}-\frac{P_{G\setminus\omega}}{d_{\omega\setminus G}}-\frac{P_{G\cup\omega}}{d_{G\cap\omega}}\Bigl).
\end{eqnarray}
One then immediately finds that that the steady-state solutions $P_G' = \lim_{t\to\infty}P_G(t)$ and $P_{G\Delta\omega}' = \lim_{t\to\infty}P_{G\Delta\omega}(t)$ of these equations yield Eq.~\eqref{eq:unitary_purity_evo} in the main text.

Let us now derive Eq.~\eqref{eq:measurement_evo} in the main text from Eq.~\eqref{eq:cont_meas_main}.  Consider the mean-field approximation for the continuous-measurement evolution in Eq.~\eqref{eq:cont_meas_main} restricted to the case where the measurement is performed on a single qudit $\Omega$:
\begin{align}\label{eq:cont_meas_one}
\frac{dP_{G}}{dt}  = \zeta \Bigl(P_{G\cup\Omega}-P_GP_{\Omega}\Bigl) \delta_{\Omega \in F \setminus G} +\zeta \Bigl(P_{(F\setminus G)\cup\Omega}-P_{F\setminus G}P_{\Omega}\Bigl) \delta_{\Omega \in G}.  
\end{align}
Let us further consider the case 
$\Omega\notin G$, i.e.\ the qudit being measured is outside of the subsystem of interest.  
The dynamics can then be evaluated by solving the following closed set of equations: 
\begin{eqnarray} \label{eq:SM_P_coupled_ode1}
\frac{d}{dt}P_{G} &=& \zeta\Bigl(P_{G\cup\Omega}-P_GP_{\Omega}\Bigl),\\ \label{eq:SM_P_coupled_ode2}
\frac{d}{dt} P_{G\cup\Omega} &=& \zeta\Bigl(P_{G}-P_{G\cup\Omega}P_{\Omega}\Bigl),\\ \label{eq:SM_P_coupled_ode3}
\frac{d}{dt} P_{\Omega} &=& \zeta\Bigl(1-P^2_{\Omega}\Bigl).
\end{eqnarray}
Here 
we used identities  $P_{F\setminus\Omega} = P_\Omega$, $P_{F\setminus(G\cup\Omega)} = P_{G\cup\Omega}$, and $P_{(F\setminus(G\cup\Omega))\cup\Omega} = P_{G}$.

As expected,  the case where qubit $\Omega$ is disentanlged (i.e.\ $P_\Omega=1$) is a fixed point of Eq.\ \eqref{eq:SM_P_coupled_ode3}. 
For general $P_\Omega(0)$ and assuming time-indepedent $\zeta$, the solution of Eq.\ \eqref{eq:SM_P_coupled_ode3} is
\be \label{eq:pomt5}
P_{\Omega}(t) = 1-\frac{2}{\mu\exp(2\zeta t)+1}, \qquad \mu = \frac{1+P_\Omega(0)}{1-P_\Omega(0)}.
\ee
By introducing new variables $S^+ = P_{G}+P_{G\cup\Omega}$ and $S^- = P_{G}-P_{G\cup\Omega}$, we can decouple the remaining two equations:
\begin{eqnarray}
\frac{dS^+}{dt} &=& \zeta(1-P_\Omega)S_+,\\
\frac{dS^-}{dt} &=& -\zeta(1+P_\Omega)S_-.
\end{eqnarray}
Solving these equations and using Eq.\ (\ref{eq:pomt5}), we find
\begin{eqnarray}
S^+(t) &=& 
S^+(0)\frac{\mu+1}{\mu+\exp(-2\zeta t)},\\
S^-(t) &=& 
S^-(0)\frac{\mu+1}{\mu+\exp(-2\zeta t)}\exp(-2\zeta t).
\end{eqnarray}
In the limit $\zeta t\to\infty$, 
weak continuous measurement must reproduce a projective measurement. Indeed,
\be
\lim_{t\to\infty}S^+(t) =  \frac{2S^+(0)}{1+P_\Omega(0)}, \qquad \lim_{t\to\infty}S^-(t) =  0,
\ee
\textrm{which implies}
\be
\lim_{t\to\infty}P_G(t) = \frac{P_G(0)+P_{G\cup\Omega}(0)}{1+P_\Omega(0)}.
\ee
which, in turn, yields Eq.~\eqref{eq:measurement_evo} in the main text. 
The solution for the case $\Omega\in G$ can be obtained in a similar way leading to the same result. We thus find that, in the mean-field approximation, the long-time limit of the evolution for the purity under a weak continuous measurement leads to the corresponding result for the projective measurement under the annealed approximation.

\end{document}